\documentclass[%
reprint,
showpacs,
bibnotes, amsmath, amssymb, aps,
prb,
floatfix
]{revtex4-1}

\usepackage{graphicx}

\begin{document}

\title{Magnetization Dynamics and Damping due to Electron-Phonon Scattering in a Ferrimagnetic Exchange Model}

\author{Alexander Baral}
\email{baral@rhrk.uni-kl.de}
\author{Svenja Vollmar}
\author{Hans Christian Schneider}
\email{hcsch@physik.uni-kl.de} 

\affiliation{Physics Department and Research Center OPTIMAS, Kaiserslautern University, P. O. Box 3049, 67663 Kaiserslautern, Germany}

\date{\today}

\begin{abstract}
We present a microscopic calculation of magnetization damping for a magnetic ``toy model.'' The magnetic system consists of itinerant carriers coupled antiferromagnetically to a dispersionless band of localized spins, and the magnetization damping is due to coupling of the itinerant carriers to a phonon bath in the presence of spin-orbit coupling. Using a mean-field approximation for the kinetic exchange model and assuming  the spin-orbit coupling to be of the Rashba form, we derive Boltzmann scattering integrals for the distributions and spin coherences in the case of an antiferromagnetic exchange splitting, including a careful analysis of the connection between lifetime broadening and the magnetic gap. For the Elliott-Yafet type itinerant spin dynamics we extract dephasing and magnetization times $T_{1}$ and $T_{2}$ from initial conditions corresponding to a tilt of the magnetization vector, and draw a comparison to phenomenological equations such as the Landau-Lifshitz (LL) or the Gilbert damping. We also analyze magnetization precession and damping for this system including an anisotropy field and find a carrier mediated dephasing of the localized spin via the mean-field coupling. 
\end{abstract}


\maketitle

\section{Introduction \label{Introduction}}
There are two widely-known phenomenological approaches to describe the damping of a precessing magnetization in an excited ferromagnet: one introduced originally by Landau and Lifshitz~\cite{LandauLifshitz} and one introduced by Gilbert,~\cite{Gilbert} which are applied to a variety of problems~\cite{Mills_LLG} involving the damping of precessing magnetic moments. Magnetization damping contributions and its inverse processes, i.e., spin torques, in particular in thin films and nanostructures, are an extremely active field, where currently the focus is on the determination of novel physical processes/mechanisms. Apart from these questions there is still a debate whether the Landau-Lifshitz or the Gilbert damping is the correct one for ``intrinsic'' damping, i.e., neglecting interlayer coupling, interface contributions, domain structures and/or eddy currents. This intrinsic damping is believed to be caused by a combination of spin-orbit coupling and scattering mechanisms such as exchange scattering between s and d electrons and/or electron-phonon scattering.~\cite{Garanin_PhysicaA91,Gerasimenko, Bauer} Without reference to the microscopic mechanism, different macroscopic analyses, based, for example, on irreversible thermodynamics or near equilibrium Langevin theory, prefer one or the other description.~\cite{Saslow, Smith} However, material parameters of typical ferromagnetic heterostructures are such that one is usually firmly in the small damping regime so that several ferromagnetic resonance (FMR) experiments were not able to detect a noticeable difference between Landau Lifshitz and Gilbert magnetization damping. A recent analysis that related the Gilbert term directly to the spin-orbit interaction arising from the Dirac equation does not seem to have conclusively solved this discussion.~\cite{Zhang_2009}

The dephasing term in the Landau-Lifshitz form is also used in models based on classical spins coupled to a bath, which  have been successfully applied to out-of-equilibrium magnetization dynamics and magnetic switching scenarios.~\cite{Nowak_Switching} The most fundamental of these are the stochastic Landau-Lifshitz equations,~\cite{Nowak_Switching, Nowak_Ultrafast, Zhang_2004, Fesenko_LLB} from which the Landau-Lifshitz Bloch equations,~\cite{Garanin_PRB97, Fesenko_Ultrafast} can be derived via a Fokker-Planck equation.

Quantum-mechanical treatments of the equilibrium magnetization in bulk ferromagnets at finite temperatures are extremely involved. The calculation of non-equilibrium magnetization phenomena and damping for quantum spin systems in more than one dimension, which include both magnetism and carrier-phonon and/or carrier-impurity interactions, at present have to employ simplified models. For instance, there have been microscopic calculations of Gilbert damping parameters based on Kohn-Sham wave functions for metallic ferromagnets~\cite{Garate_PRB2009,Gilmore_PRB2011} and Kohn-Luttinger $p$-$d$ Hamiltonians for magnetic semiconductors.~\cite{Sinova} While the former approach uses spin density-functional theory, the latter approach treats the anti-ferromagnetic kinetic-exchange coupling between itinerant $p$-like holes and localized magnetic moments originating from impurity $d$-electrons within a mean-field theory. In both cases, a constant spin and band-independent lifetime for the itinerant carriers is used as an input, and a Gilbert damping constant is extracted by comparing the quantum mechanical result for $\omega\to 0$ with the classical formulation. There have also been investigations, which extract the Gilbert damping for magnetic semiconductors from a microscopic calculation of carrier dynamics including Boltzmann-type scattering integrals.~\cite{Shen_2010,Shen_2012} Such a kinetic approach, which is of a similar type as the one we present in this paper, avoids the introduction of electronic lifetimes because the scattering is calculated dynamically.

The present paper takes up the question how the spin dynamics in the framework of the macroscopic Gilbert or Landau-Lifshitz damping compare to a microscopic model of relaxation processes in the framework of a relatively simple model. We analyze a mean-field kinetic exchange model including spin-orbit coupling for the itinerant carriers. Thus the magnetic mean-field dynamics is combined with a microscopic description of damping provided by the electron-phonon coupling. This interaction transfers energy and angular momentum from the itinerant carriers to the lattice. The electron-phonon scattering is responsible both for the lifetimes of the itinerant carriers and the magnetization dephasing. The latter occurs because of spin-orbit coupling in the states that are connected by electron-phonon scattering. To be more specific, we choose an anti-ferromagnetic coupling at the mean-field level between itinerant electrons and a dispersion-less band of localized spins for the magnetic system. To keep the analysis simple we use as a model for the spin-orbit coupled itinerant carrier states a two-band Rashba model. As such it is a single-band version of the multi-band Hamiltonians used for III-Mn-V ferromagnetic semiconductors.~\cite{MacDonald_Review,Dietl_Review,MacDonald_Curie,MacDonald_SpinWave,Sinova} The model analyzed here also captures some properties of two-sublattice ferrimagnets, which are nowadays investigated because of their magnetic switching dynamics.~\cite{Rasing,Alebrand} The present paper is set apart from studies of spin dynamics in similar models with more complicated itinerant band structures~\cite{Shen_2010,Shen_2012} by a detailed comparison of the phenomenological damping expressions with a microscopic calculation as well as a careful analysis of the restrictions placed by the size of the magnetic gap on the single-particle broadening in Boltzmann scattering.

This paper is organized as follows. As an extended introduction, we review in Sec.~\ref{DefDephasing} some basic facts concerning the Landau-Lifshitz and Gilbert damping terms on the one hand and the Bloch equations on the other. In Sec.~\ref{LLvsBloch} we point out how these different descriptions are related in special cases. We then introduce a microscopic model for the dephasing due to electron-phonon interaction in Sec.~\ref{micro_model}, and present numerical solutions for two different scenarios in Secs.~\ref{simulation_iso} and \ref{simulation_aniso}. The first scenario is the dephasing between two spin subsystems (Sec.~\ref{simulation_iso}), and the second scenario is a relaxation process of the magnetization toward an easy-axis (Sec.~\ref{simulation_aniso}). A brief conclusion is given at the end.

\section{Phenomenologic Descriptions of Dephasing and Relaxation \label{DefDephasing}}

\begin{figure}[tb]
\centering
\includegraphics[width=0.45\textwidth]{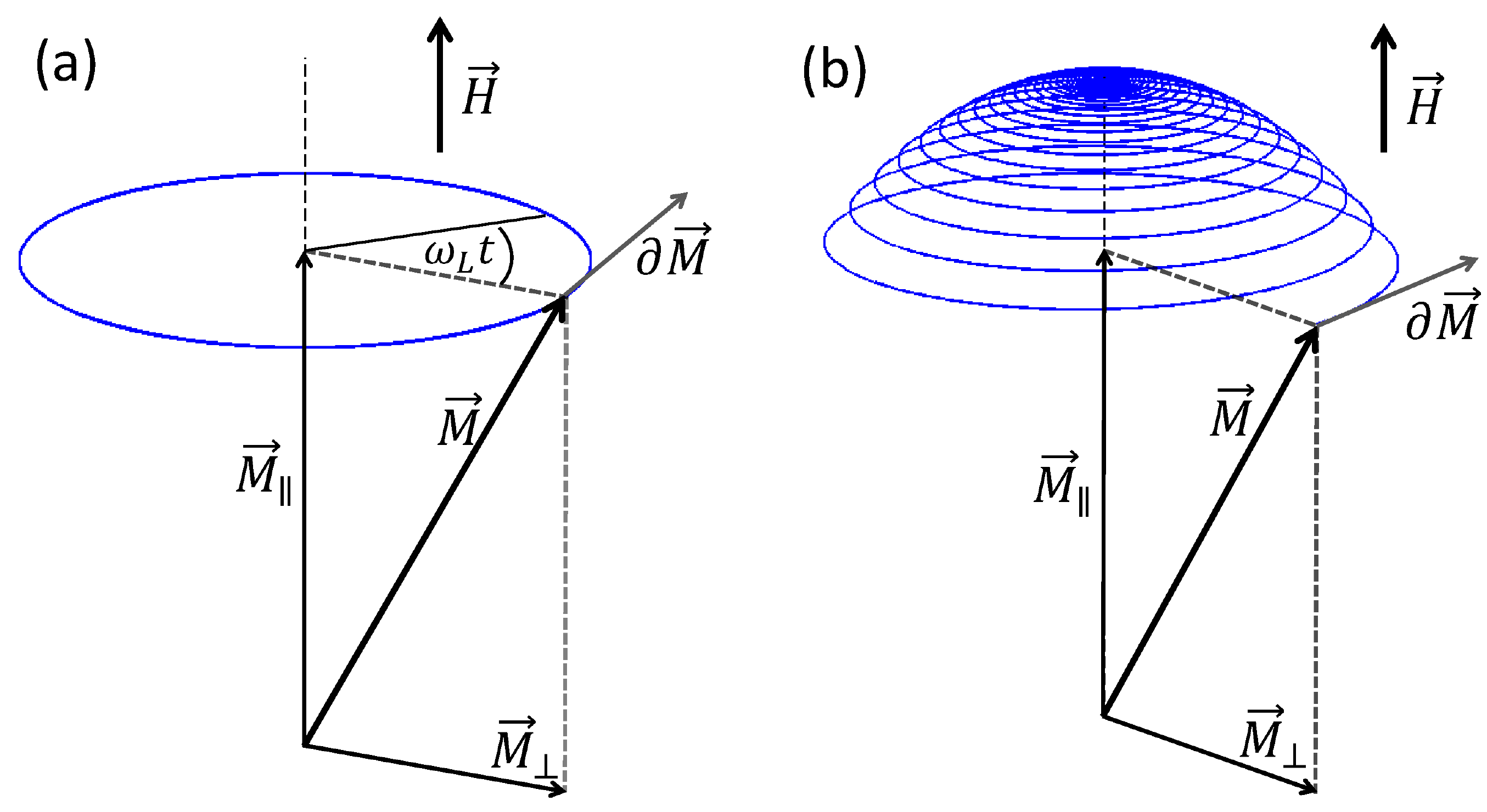}
\caption{Illustration of non-equilibrium spin-dynamics in presence of a magnetic field without relaxation (a) and within relaxation (b).}
\label{fig:Illus_Deph}
\end{figure}

We summarize here some results pertaining to a single-domain ferromagnet, and set up our notation. In equilibrium we assume the magnetization to be oriented along its easy axis or a magnetic field $\vec{H}$, which we take to be the $z$ axis in the following. If the magnetization is tilted out of equilibrium, it starts to precess. As illustrated in Fig.~\ref{fig:Illus_Deph} one distinguishes the longitudinal component $M_{\|}$, in $z$ direction, and the transverse part $M_{\perp} \equiv \sqrt{M^2 - M_{\parallel}^2}$, precessing in the $x$-$y$ plane with the Larmor frequency $\omega_{\mathrm{L}}$.

In connection with the interaction processes that return the system to equilibrium, the decay of the transverse component is called dephasing.
There are three phenomenological equations used to describe spin dephasing processes:
\begin{enumerate}
	\item The Bloch(-Bloembergen) equations~\cite{Bloch, Bloembergen}
	\begin{align}
		\label{Bloch_eq_long}
		\frac{\partial}{\partial t} M_{\parallel}(t) &= - \frac{M_{\parallel}(t) - M^{\mathrm{eq}}}{T_{1}} \\
		\label{Bloch_eq_perp}
		\frac{\partial}{\partial t} M_{\perp}(t) &= - \frac{M_{\perp}(t)}{T_{2}}
	\end{align}
describe an exponential decay towards the equilibrium magnetization $M^{\mathrm{eq}}$ in $z$ direction. The transverse component decays with a time constant $T_{2}$, whereas the longitudinal component approaches its equilibrium amplitude with $T_{1}$.
 These time constants may be fit independently to experimental results or microscopic calculations. 	
\item  Landau-Lifshitz damping~\cite{LandauLifshitz} with parameter $\lambda$
	\begin{equation}
		\label{LL_eq}
		\frac{\partial}{\partial t} \vec{M}(t) = - \gamma \vec{M} \times \vec{H} - \lambda \ \frac{\vec{M}}{M} \times \big(\vec{M} \times \vec{H}\big)
	\end{equation}
where $\gamma$ is the gyromagnetic ratio. The first term models the precession with a frequency $\omega_{\mathrm{L}} = \gamma |\vec{H}|$, whereas the second term is solely responsible for damping.	
\item Gilbert damping~\cite{Gilbert} with the dimensionless Gilbert damping parameter~$\alpha$
	\begin{equation}
		\label{Gilbert_eq}
		\frac{\partial}{\partial t} \vec{M}(t) = - \gamma_{\mathrm{G}} \vec{M} \times \vec{H} + \alpha  \Big(\frac{\vec{M}}{M} \times 				\partial_{t} \vec{M}\Big)  
	\end{equation}
It is generally accepted that $\alpha$ is independent of the static magnetic fields $\vec{H}$ such as anisotropy fields,~\cite{Sinova, Vonsovskii} and thus depends only on the material and the microscopic interaction processes. 
\end{enumerate}

The Landau-Lifshitz and Gilbert forms of damping are mathematically equivalent~\cite{Gilbert,Stancil,Saslow} with
\begin{align}
		\alpha &= \frac{\lambda}{\gamma}
		\label{alpha_lambda_relation} \\
		\gamma_G &= \gamma (1 + \alpha^2)
		\label{gammaG_gamma_relation}	
\end{align}
but there are important differences. In particular, an increase of $\alpha$ lowers the precession frequency in the dynamics with Gilbert damping, while the damping parameter $\lambda$ in the Landau-Lifshitz equation has no impact on the precession. In contrast to the Bloch equations, Landau-Lifshitz and Gilbert spin-dynamics always conserve the length $|\vec{M}|$ of the magnetization vector. 

An argument by Pines and Slichter,~\cite{Slichter} shows that there are two different regimes for Bloch-type spin dynamics depending on the relation between the Larmor period and the correlation time. As long as the correlation time is much longer than the Larmor period, the system ``knows'' the direction of the field during the scattering process. Stated differently, the scattering process ``sees'' the magnetic gap in the bandstructure. Thus, transverse and longitudinal spin components are distinguishable and the Bloch decay times $T_{1}$ and $T_{2}$ can differ. If the correlation time is considerably shorter than the Larmor period, this distinction is not possible, with the consequence that $T_1$ must be equal to $T_2$. Within the microscopic approach, presented in Sec.~\ref{eph_scattering}, this consideration shows up again, albeit for the energy conserving $\delta$ functions resulting from a Markov approximation. 

The regime of short correlation times has already been investigated in the framework of a microscopic calculation by Wu and coworkers.~\cite{Wu} They analyze the case of a moderate external magnetic field applied to a non-magnetic n-type GaAs quantum well and include different scattering mechanisms (electron-electron Coulomb, electron-phonon, electron-impurity). They argue that the momentum relaxation rate is the crucial time scale in this scenario, which turns out to be much larger than the Larmor frequency. Their numerical results confirm the identity $T_{1} = T_{2}$ expected from the Pines-Slichter argument.

\section{Relation between Landau-Lifshitz, Gilbert and Bloch \label{LLvsBloch}}

We highlight here a connection between the Bloch equations (\ref{Bloch_eq_long}, \ref{Bloch_eq_perp}) and the Landau-Lifshitz equation~\eqref{LL_eq}. To this end we assume a small initial tilt of the magnetization and describe the subsequent dynamics of the magnetization in the form
	\begin{equation}
		\label{LL_approach}
		\vec{M}(t) = \begin{pmatrix} \delta M_{\perp}(t) \cos(\omega_{\mathrm{L}} t) \\ \delta M_{\perp}(t) \sin(\omega_{\mathrm{L}} t) \\ M^{\mathrm{eq}} - \delta M_{\parallel}(t) \end{pmatrix}
	\end{equation}
where $\delta M_{\perp}$ and $\delta M_{||}$ describe deviations from equilibrium. Putting this into eq.~\eqref{LL_eq} one gets a coupled set of equations.
\begin{align}
		\label{LL_coupled}
					\frac{\partial}{\partial t} \delta M_{\perp}(t) &= - \lambda H \frac{M^{\mathrm{eq}} - \delta M_{\parallel}(t) }{ |\vec{M}(t)| } \delta M_{\perp}(t) \\
			\frac{\partial}{\partial t} \delta M_{\parallel}(t) &= - \lambda H  \frac{1}{ |\vec{M}(t)| }  \delta M^{2}_{\perp}(t)
		\end{align}
Eq.~\eqref{LL_coupled} is simplified for a small deviation from equilibrium, i.e.,  $\delta M(t) \ll M^{\mathrm{eq}}$ and $|\vec{M}(t)| \approx M^{\mathrm{eq}}$:
	\begin{align}
		\label{LL_solution}
			\delta M_{\perp}(t) &= C \ \exp( - \lambda H t) \\
			\delta M_{\parallel}(t) &= \frac{C^2}{2 M^{\mathrm{eq}}} \exp( - 2 \lambda H t)
	\end{align}
where $C$ is an integration constant. For small excitations the deviations decay exponentially and Bloch decay times $T_{1}$ and $T_{2}$  result, which are related by
	\begin{equation}
		\label{T12_lambda_relation}
		2 T_{1} = T_{2} = \frac{1}{\lambda H}.
	\end{equation}
	Only this ratio of the Bloch times is compatible with a constant length of the magnetization vector at low excitations. By combining Eqs.~\eqref{T12_lambda_relation} and~\eqref{alpha_lambda_relation} one can connect the Gilbert parameter $\alpha$ and the dephasing time~$T_{2}$	
	\begin{equation}
		\label{alpha_T2_relation}
		\alpha = \frac{1}{T_{2} \ \omega_{\mathrm{L}}}.
	\end{equation}
	If the conditions for the above approximations apply, the Gilbert damping parameter $\alpha$ can be determined by fitting the dephasing time $T_{2}$ and the Larmor frequency $\omega_{\mathrm{L}}$ to computed or measured spin dynamics. This dimensionless quantity is well suited to compare the dephasing that results from different relaxation processes.

	\begin{figure}[tb]
		\centering
			\includegraphics[width=0.5\textwidth]{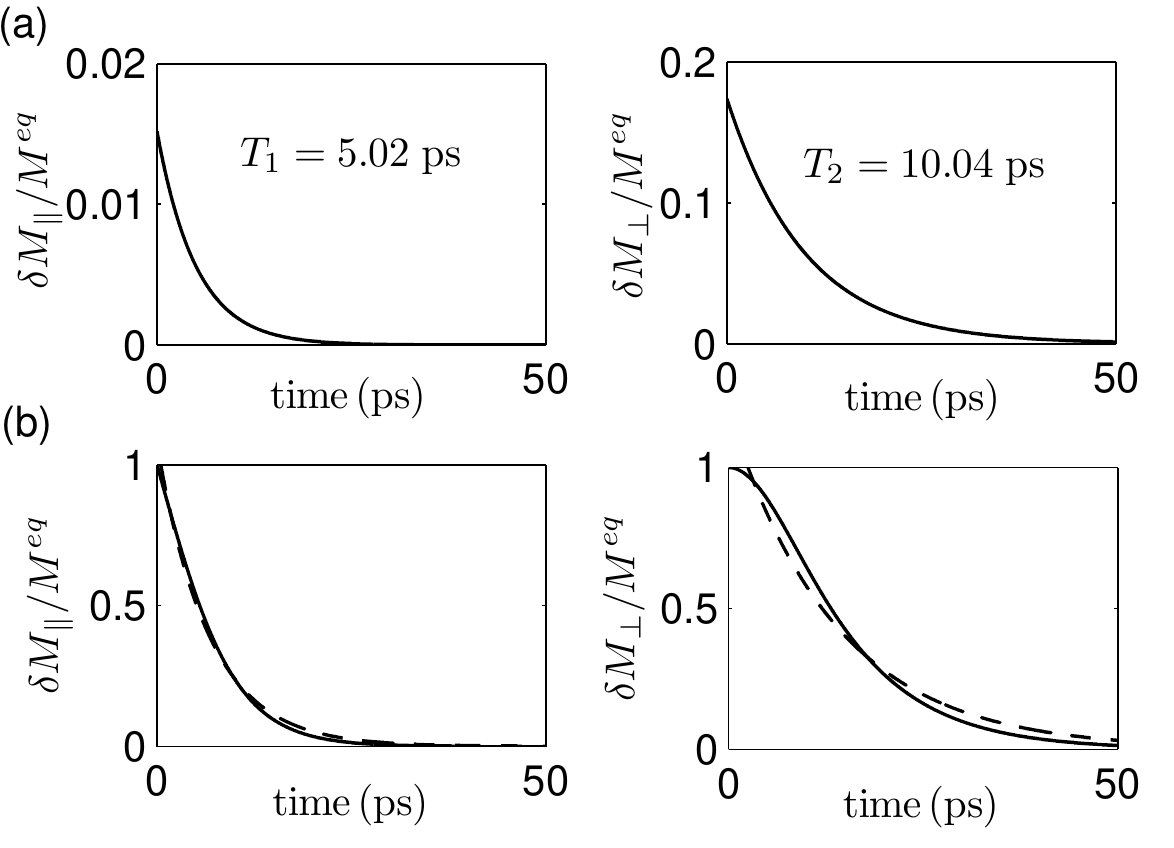}
			\caption{Dynamics of $\delta M_{\perp}$ and $\delta M_{\parallel}$ computed using to Landau-Lifshitz damping ($\omega_{\mathrm{L}} = 1 \, \mathrm{ps^{-1}}$, $H = 10^{6} \, \mathrm{\frac{A}{m}} \approx 1.26 \cdot 10^{4} \, \mathrm{Oe}$, $\lambda = 10^{-7} \, \mathrm{\frac{m}{A \, ps}}$). (a) An angle of $10^\circ$ leads to exponential an exponential decay with well defined $T_{1}$ and $T_{2}$ times. (b). For an angle of $90^\circ$, the decay (solid line) is not exponential as comparison with the exponential fit (dashed line) clearly shows.}
		\label{fig:LL_fig}
	\end{figure}
	
Figure~\ref{fig:LL_fig} shows the typical magnetization dynamics that results from \eqref{LL_eq}, i.e., Landau-Lifshitz damping. As an illustration of a small excitation we choose in Fig.~\ref{fig:LL_fig}(a) an angle of $10^{\circ}$ for the initial tilt of the magnetization, which results in an exponential decay with $2 T_{1} = T_{2}$. From the form of Eq.~\eqref{LL_eq} it is clear that this behavior persists even for large $\omega_{\mathrm{L}}$ and $\lambda$. Obviously the Landau-Lifshitz and Gilbert damping terms describe a scenario with relatively long correlation times (i.e., small scattering rates), because only in this regime both decay times can differ. The microscopic formalism in Sec.~\ref{micro_model} works in the same regime and will be compared with the phenomenological results. For an excitation angle of $90^{\circ}$, the Landau-Lifshitz dynamics shown in Fig.~\ref{fig:LL_fig}(b) become non-exponential, so that no well-defined Bloch decay times $T_1$, $T_2$ exist.

\section{Microscopic Model \label{micro_model}}

In this section we describe a microscopic model that includes magnetism at the mean-field level, spin-orbit coupling as well as the microscopic coupling to a phonon bath treated at the level of Boltzmann scattering integrals. We then compare the microscopic dynamics to the Bloch equations~\eqref{Bloch_eq_long}, \eqref{Bloch_eq_perp}, as well as the Landau-Lifshitz~\eqref{LL_eq} and Gilbert damping terms~\eqref{Gilbert_eq}. The magnetic properties of the model are defined by an antiferromagnetic coupling between localized magnetic impurities and itinerant carriers. As a prototypical spin-orbit coupling we consider an effectively two-dimensional model with a Rashba spin-orbit coupling. The reason for the choice of a model with a two-dimensional wave vector space is not an investigation of magnetization dynamics with reduced dimensionality, but rather a reduction in the dimension of the integrals that have to be solved numerically in the Boltzmann scattering terms. Since we treat the exchange between the localized and itinerant states in a mean-field approximation, our two-dimensional model still has a ``magnetic ground state'' and presents a framework, for which qualitatively different approaches can be compared. We do not aim at quantitative predictions for, say, magnetic semiconductors or ferrimagnets with two sublattices.  Finally, we include a standard interaction hamiltonian between the itinerant carriers and acoustic phonons. The corresponding hamiltonian reads
\begin{equation}
	\label{Entire_Ham}
	\hat{\mathcal{H}} = \hat{\mathcal{H}}_{\mathrm{mf}} + \hat{\mathcal{H}}_{\mathrm{so}} + \hat{\mathcal{H}}_{\mathrm{e-ph}} + \hat{\mathcal{H}}_{\mathrm{aniso}} .
\end{equation}
Only in Sec.~\ref{simulation_aniso} an additional field  $\hat{\mathcal{H}}_{\mathrm{aniso}}$ is included, which is intended to model a small anisotropy.
%
%
%
\subsection{Exchange interaction between itinerant carriers and localized spins \label{toymodel}}

\begin{figure}[tb]
	\centering
		\includegraphics[width=0.45\textwidth]{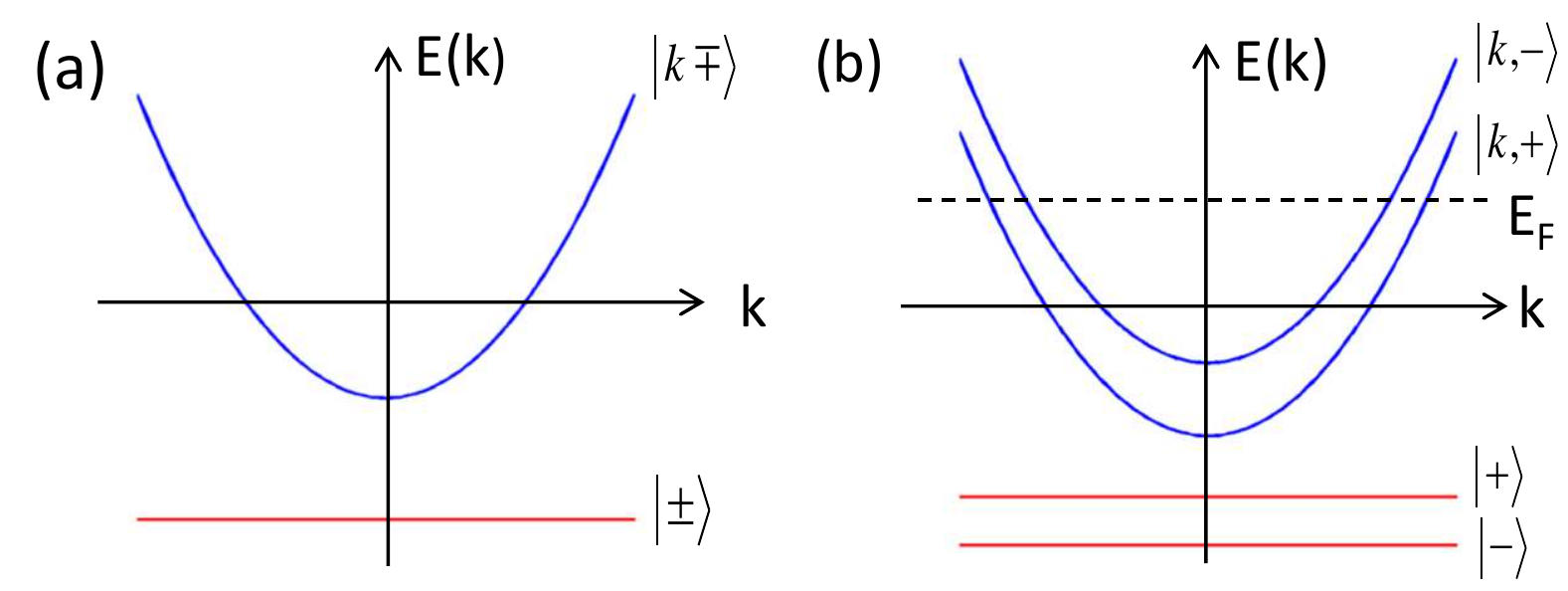}
	\caption{Sketch of the band-structure with localized (flat dispersions) and itinerant (parabolic dispersions) electrons. Above the Curie-Temperature $T_{\mathrm{C}}$ the spin-eigenstates are degenerate (a), whereas below $T_{\mathrm{C}}$ a gap between the spin states exists.}
	\label{fig:Kondo_bandstructure}
\end{figure}

The ``magnetic part'' of the model is described by the Hamiltonian
\begin{equation}
	\label{KondoHam}
	\hat{\mathcal{H}}_{\mathrm{mf}} = \sum_{\vec{k} \mu} \frac{\hbar^2 k^2}{2 m^{*}} \hat{c}^\dagger_{\vec{k} \mu} \hat{c}_{\vec{k} \mu} + J \, \hat{\vec{s}} \cdot \hat{\vec{S}} .
\end{equation}
which we consider in the mean-field limit. The first term represents itinerant carriers with a $k$-dependent dispersion relation. In the following we assume $s$-like wave functions and parabolic energy dispersions. The effective mass is chosen to be $m^{*} = 0.5 \, m_{e}$, where $m_{e}$ is the free electron mass, and the $\hat{c}^{(\dagger)}_{\vec{k} \mu}$ operators create and annihilate carriers in the state $| \vec{k}, \mu \rangle$ where $\mu$ labels the itinerant bands, as shown in Fig.~\ref{fig:Kondo_bandstructure}(a).

The second term describes the coupling between itinerant spins $\vec{s}$ and localized spins $\vec{S}$ via an antiferromagnetic exchange interaction
\begin{align}
	\label{def-itinSpin}
	\hat{\vec{s}} &= \frac{1}{2} \sum_{\vec{k}} \sum_{\mu \mu'} \, \langle \vec{k}, \mu' |\hat{\vec{\sigma}}| \vec{k}, \mu \rangle \hat{c}^\dagger_{\vec{k} \mu} \hat{c}_{\vec{k} \mu'} \\
		\label{def-locSpin}
	\hat{\vec{S}} &= \frac{1}{2} \sum_{\nu \nu'} \, \langle  \nu' |\hat{\vec{\sigma}}| \nu \rangle \sum_{\vec{K}} \hat{C}^{\dagger}_{\vec{K} \nu} \hat{C}_{\vec{K} \nu'}
\end{align}
Here, we have assumed that the wave functions of the localized spins form dispersionless bands, i.e., we have implicitly introduced a virtual-crystal approximation. Due to the assumption of strong localization there is no orbital overlap between these electrons, which are therefore considered to have momentum independent eigenstates $ \left| \nu \right. \rangle $ and a flat dispersion, as illustrated in Fig.~\ref{fig:Kondo_bandstructure}(a). The components of the vector $\hat{\vec{\sigma}}$ are the Pauli matrices $\hat{\sigma}_{i}$ with $i=x,y,z$, and  $\hat{C}_{\vec{K}\nu}^{(\dagger)}$ are the creation and annihilation operators for a localized spin state. 

We do \emph{not} include interactions among localized or itinerant spins, such as exchange scattering. For simplicity, we assume both itinerant and localized electrons to have a spin $1/2$ and therefore $\mu$ and $\nu$ to run over two spin-projection quantum numbers $\pm 1/2$. In the following we chosse an antiferromagnetic ($J > 0$) exchange constant $J= 500\,\mathrm{meV}$, which leads to the schematic band structure shown in Fig.~\ref{fig:Kondo_bandstructure}(b).

In the mean field approximation used here, the itinerant carriers feel an effective magnetic field $\hat{H}_{\mathrm{loc}}$ 
\begin{equation}
	\label{eff_HField}
	\vec{H}_{\mathrm{loc}} = - \frac{J \mu_{\mathrm{B}} \mu}{g}  \, \vec{S}
\end{equation}
caused by localized moments and vice versa. Here $\mu_{\mathrm{B}}$ is the Bohr magneton and $g=2$ is the g-factor of the electron. The permeability $\mu$ is assumed to be the vacuum permeability $\mu_{0}$. This time-dependent magnetic field $\vec{H}_{\mathrm{loc}}(t)$ defines the preferred direction in the itinerant sub-system and therefore determines the longitudinal and transverse component of the itinerant spin at each time.

\subsection{Rashba spin-orbit interaction\label{Rashba}}

The Rashba spin-orbit coupling is given by the Hamiltonian
\begin{equation}
	\label{RashbaHam}
	\hat{\mathcal{H}}_{\mathrm{so}} = \alpha_{\mathrm{R}} \left( \hat{\sigma}_{x} k_{y} - \hat{\sigma}_{y} k_{x} \right) \\
\end{equation}
A Rashba coefficient of  $\alpha_{\mathrm{R}} = 10 \, \mathrm{meV\,nm}$ typical for semiconductors is chosen in the following calculations. This value, which is close to the experimental one for the InSb/InAlSb material system,~\cite{Leontiadou_2011} is small compared to the exchange interactions, but it allows the exchange of angular momentum with the lattice.
%
%
\subsection{Coherent dynamics \label{Basis}} 

From the above contributions \eqref{KondoHam} and \eqref{RashbaHam} to the Hamiltonian we derive the equations of motion containing the coherent dynamics due to the exchange interaction and Rashba spin-orbit coupling as well as the incoherent electron-phonon scattering. We first focus on the coherent contributions. In principle, one has the choice to work in a basis with a fixed spin-quantization axis or to use single-particle states that diagonalize the mean-field (plus Rashba) Hamiltonian. Since we intend to use a Boltzmann scattering integral in Sec.~\ref{eph_scattering} we need to apply a Markov approximation, which only works if one deals with diagonalized eigenenergies. In our case this is the single-particle basis that diagonalizes the entire one-particle contribution of the Hamiltonian $\hat{\mathcal{H}}_{\mathrm{mf}} + \hat{\mathcal{H}}_{\mathrm{so}}$. In matrix representation this one-particle contribution for the itinerant carriers reads:
\begin{equation}
\label{OneParticleHam}
\hat{\mathcal{H}}_{\mathrm{mf}} + \hat{H}_{\mathrm{so}} =
	\begin{pmatrix}	
		\frac{\hbar^2 k^2}{2 m^{*}} + \Delta^{\mathrm{loc}}_{z} & ( \Delta^{\mathrm{loc}}_{+} + R_{\vec{k}} )^{*} \\	
		\Delta^{\mathrm{loc}}_{+} + R_{\vec{k}} 				& \frac{\hbar^2 k^2}{2 m^{*}} - \Delta^{\mathrm{loc}}_{z}
	\end{pmatrix} 
\end{equation}
where we have defined $\Delta^{\mathrm{loc}}_{i} = J \frac{1}{2} \langle\hat{S}_{i}\rangle$ and  $R_{\vec{k}} = -i \alpha_{\mathrm{R}} k \exp(i \varphi_{k})$ with $\varphi_{k} = \arctan (k_y/k_x)$. The eigenenergies are
\begin{equation}
	\label{itin_eigenenergy}
	\epsilon_{\vec{k}}^{\pm} = \frac{\hbar^2 k^2}{2 m^{*}} \mp \sqrt{ |\Delta_{z}^{\mathrm{loc}}|^2 + |R_{\vec{k}} + \Delta_{+}^{\mathrm{loc}}|^2 }.
\end{equation} 
and the eigenstates
\begin{equation}
	\label{itin_eigenstates}
	| \vec{k}, + \rangle = \begin{pmatrix} 1 \\ \xi_{\vec{k}} \end{pmatrix}  \ ; \  | \vec{k}, - \rangle = \begin{pmatrix} -\xi_{\vec{k}}^{*} \\ 1 \end{pmatrix}
\end{equation}
where
\begin{equation}
	\label{itin_xi_def}
	\xi_{\vec{k}} = \frac{ \Delta^{\mathrm{loc}}_{+} + R_{\vec{k}} }{\Delta^{\mathrm{loc}}_{z} + \sqrt{ |\vec{\Delta}^{\mathrm{loc}}|^2 + |R_{\vec{k}}|^2} }
\end{equation}

In this basis  the coherent part of the equation of motion for the itinerant density matrix $\rho_{\vec{k}}^{\mu \mu'} \equiv \langle \hat{c}^\dagger_{\vec{k} \mu} \hat{c}_{\vec{k} \mu'} \rangle$ reads
\begin{equation}
	\label{itinEoM_KondoRashba}
	\frac{\partial}{\partial t} \rho_{\vec{k}}^{\mu \mu'}\Big|_{\mathrm{coh}} = \frac{i}{\hbar} \big( \epsilon_{\vec{k}}^{\mu} - \epsilon_{\vec{k}}^{\mu'} \big)  \rho_{\vec{k}}^{\mu \mu'}.
\end{equation}
No mean-field or Rashba terms appear explicitly in these equations of motion since their contributions are now hidden in the time-dependent eigenstates and eigenenergies. 
Since we are interested in dephasing and precessional dynamics, we assume a comparatively small spin-orbit coupling, that can dissipate angular momentum into the lattice, but does not have a decisive effect on the band-structure. Therefore we use the spin-mixing only in the transition matrix elements of the electron-phonon scattering $M_{\vec{k} \mu}^{\vec{k}' \mu'}$~\eqref{phMatElem_eq}. For all other purposes we set $R_{\vec{k}} = 0$. In particular, the energy-dispersion $\epsilon_{\vec{k}}^{\pm}$ is assumed to be unaffected by the spin-orbit interaction and therefore it is spherically symmetric. 

With this approximation the itinerant eigenstates are always exactly aligned with the effective field of the localized moments $\vec{H}_{\mathrm{loc}}(t)$. Since this effective field changes with time, the diagonalization and a transformation of the spin-density matrix in ``spin space'' has to be repeated at each time-step. This effort makes it easier to identify the longitudinal and transverse spin components with the elements of the single-particle density matrix: The off-diagonal entries of the density matrix $\rho_{\vec{k}}^{\pm \mp}$, which precess with the $k$-independent Larmor frequency $\omega_{\mathrm{L}} =  2 \Delta^{\mathrm{loc}} / \hbar$, always describe the dynamics of the transverse spin-component. The longitudinal component, which does not precess, is represented by the diagonal entries $\rho_{\vec{k}}^{\pm \pm}$. Since both components change their spatial orientation continuously, we call this the rotating frame. The components of the spin vector in the rotating frame are
\begin{align}
	\langle \hat{s}_{\parallel} \rangle &= \frac{1}{2} \sum_{\vec{k}} \big( \rho_{\vec{k}}^{+ +} - \rho_{\vec{k}}^{- -} \big)
	\label{rotating-frame-parallel} \\
	\langle \hat{s}_{\perp} \rangle &= \sum_{\vec{k}} \big|\rho_{\vec{k}}^{+ -}\big|
	\label{rotating-frame-perp}
\end{align}
The components in the fixed frame are obtained from Eq.~\eqref{def-itinSpin} 
\begin{equation}
	\label{fixed frame}
	\langle \hat{\vec{s}} \rangle = \frac{1}{2} \sum_{\vec{k}} \sum_{\mu \mu'} \, \langle \vec{k}, \mu'|\hat{\vec{\sigma}}| \vec{k}, \mu \rangle \ \rho_{\vec{k}}^{\mu \mu'}
\end{equation}
In this form, the time-dependent states carry the information how the spatial components are described by the density matrix at each time step. No time-independent ``longitudinal'' and ``transverse'' directions can be identified in the fixed frame. 

In a similar fashion, the diagonalized single-particle states of the localized spin system are obtained. The eigenenergies are
\begin{equation}
	\label{loc_eigenenergy}
	E^{\pm} =  \mp \big|\vec{\Delta}^{\mathrm{itin}}\big|
\end{equation} 
where $\Delta^{\mathrm{itin}}_{i} = J \frac{1}{2} \langle \hat{s}_{i} \rangle$ is the localized energy shift caused by the itinerant spin component $s_{i}$. The eigenstates are again always aligned with the itinerant magnetic moment. In this basis the equation of motion of the localized spin-density matrix $ \rho_{\mathrm{loc}}^{\nu \nu'} \equiv \sum_{\vec{K}} \langle \hat{C}^{\dagger}_{\vec{K} \nu} \hat{C}_{\vec{K} \nu'} \rangle$ is simply \begin{equation}
	\label{locEoM_Kondo}
	\frac{\partial}{\partial t} \rho_{\mathrm{loc}}^{\nu \nu'} = \frac{i}{\hbar} ( E^{\nu} - E^{\nu'} )  \rho_{\mathrm{loc}}^{\nu \nu'}
\end{equation}
and does not contain explicit exchange contributions. Eqs.~\eqref{rotating-frame-parallel}, \eqref{rotating-frame-perp}, and \eqref{fixed frame} apply in turn to the components $\langle S_\|\rangle$ and $\langle S_\perp\rangle$ of the localized spin and its spin-density matrix $\rho_{\mathrm{loc}}^{\nu \nu'}$.

\subsection{Electron-phonon Boltzmann scattering with spin splitting\label{eph_scattering}}

Relaxation is introduced into the model by the interaction of the itinerant carriers with a phonon bath, which plays the role of an energy and angular momentum sink for these carriers. Our goal here is to present a derivation of the Boltzmann scattering contributions using standard methods, see, e.g., Refs.~\onlinecite{HaugKoch,Raichev}. However, we emphasize that describing interaction as a Boltzmann-like instantaneous, energy conserving scattering process is limited by the existence of the magnetic gap. Since we keep the spin mixing due to Rashba spin-orbit coupling only in the Boltzmann scattering integrals, the resulting dynamical equations describe an Elliott-Yafet type spin relaxation.

The  electron-phonon interaction Hamiltonian reads~\cite{HaugKoch}
\begin{equation}
	\label{ephHam}
	\begin{aligned}
		\hat{\mathcal{H}}_{\mathrm{e-ph}} =& \sum_{\vec{q}} \hbar \omega^{\mathrm{ph}}_{q} \, \hat{b}^\dagger_{\vec{q}} \hat{b}_{\vec{q}} \\
		&+  \sum_{\vec{k} \vec{k}'} \sum_{\mu \mu'} \big( M_{\vec{k} \mu}^{\vec{k}' \mu'} \hat{c}^\dagger_{\vec{k} \mu} \hat{b}_{\vec{k}-\vec{k}'} \hat{c}_{\vec{k}' \mu'}  + \text{h.c.} \big) 
	\end{aligned}
\end{equation}
where $\hat{b}^{(\dagger)}_{\vec{q}}$ are the bosonic operators, that create or annihilate acoustic phonons with momentum $\vec{q}$ and linear dispersion $\omega_{\mathrm{ph}}(q) = c_{\mathrm{ph}} |\vec{q}|$. The sound velocity is taken to be $c_{\mathrm{ph}} = 40$\,nm/ps and we use an effectively two-dimensional transition matrix element~\cite{Jauho_Phonon}
\begin{equation}
	\label{phMatElem_eq}
		 M_{\vec{k} \mu}^{\vec{k}' \mu'} =  D \, \sqrt{ |\vec{k}-\vec{k}'| } \ \langle \vec{k}, \mu | \vec{k}', \mu' \rangle
\end{equation}
where the deformation potential is chosen to be $D=60 \, \mathrm{meV}\mathrm{nm}^{1/2}$. The scalar-product between the initial state $| \vec{k}', \mu' \rangle$ and the final state $| \vec{k}, \mu \rangle$ of an electronic transition takes the spin-mixing due to Rashba spin-orbit coupling into account. 

The derivation of Boltzmann scattering integrals for the itinerant spin-density matrix \eqref{itinEoM_KondoRashba} leads to a memory integral of the following shape\begin{equation}
	\label{memory_integral}
	\frac{\partial}{\partial t} \rho_{j}(t)\Big|_{\text{inc}} = \frac{1}{\hbar} \sum_{j'} \int_{-\infty}^{t} e^{i(\Delta E_{j j'} + i \gamma)(t-t')} \, F_{j j'}[\rho(t')] \, dt',
\end{equation}
regardless whether one uses Green's function~\cite{Raichev} or equation-of-motion techniques.~\cite{HaugKoch} Since we go through a standard derivation here, we highlight only the important parts for the present case and do not write the equations out completely. In particular, for scattering process $j'=|\mu',\vec k'\rangle \rightarrow j=|\mu,\vec k \rangle $, we use $F_{j j'}[\rho(t')]$ as an abbreviation for a product of dynamical electronic spin-density matrix elements $\rho$, evaluated at time $t'<t$, and equilibrium phononic distributions. 
The corresponding energy difference is denoted by $\Delta E_{j j'} = E_{j} - E_{j'} \pm \hbar \omega_{\mathrm{ph}}(|\vec k -\vec k'|)$, and $\gamma$ describes the decay of the exponential function due to dissipation and/or higher order correlation functions. 
In general, the integral has to be evaluated numerically and contains memory effects.  To apply the Markov approximation one needs to compare two time scales: the ``memory depth'' $1/\gamma$, i.e., the time scale on which the $\exp[-\gamma (t-t')]$ factor essentially cuts off the integral, and the typical time scale on which the $F$ term changes. In this paper we deal with relaxation processes not too far away from equilibrium, so that the typical time scale of the components of the spin-density matrix contained in $F$ is set by the Bloch times $T_{1}$ and $T_{2}$. We can thus approximate $F_{j j'}[\rho(t')]$ by $F_{j j'}[\rho(t)]$, for all transitions labeled by $j$ and $j'$, if the memory depth is shorter than the Bloch time(s), or
\begin{equation}
	\label{markov_approximation}
	\gamma \gg \frac{1}{T_{1}} .
\end{equation}
Provided condition~\eqref{markov_approximation} holds, the integral~\eqref{memory_integral} can be done using the Markov approximation $F_{j j'}[\rho(t')]\simeq F_{j j'}[\rho(t)]$
\begin{equation}
	\label{energy_denominator}
	\frac{\partial}{\partial t} \rho_{j}(t)\Big|_{\mathrm{incoh}} = \frac{i}{\hbar} \sum_{j'} F_{j j'}[\rho(t)] \, \frac{1}{\Delta E_{j j'} + i \hbar \gamma}.
\end{equation}
As it is customary, we neglect in the following the real part of the complex energy denominator, which results in shifts of the single-particle energies. While these shifts may play an important role in non-Markovian problems with discrete energy levels~\cite{SchneiderChowKoch},  the imaginary parts yield the relaxation contributions that are important for the present paper
\begin{equation}
	\label{energy_Lorentz}
	\frac{\partial}{\partial t} \rho_{j}(t)\Big|_{\mathrm{incoh}} = \sum_{j'} F_{j j'}[\rho(t)] \, \frac{\hbar \gamma}{\left(\Delta E_{j j'}\right)^2 + \left(\hbar \gamma\right)^2} 
\end{equation}
All transitions are thus weighted by a Lorentzian peaked at resonant transitions ($\Delta E_{j j'} = 0$) with a broadening of $\hbar \gamma$ that may be interpreted as an energy uncertainty. For relaxation processes in a system with a spin-splitting (due to internal fields and/or spin-orbit coupling), this broadening must not be so large as to blur the distinction between the split bands. Consequently, only if the broadening $\gamma$ is smaller than the magnetic splitting, i.e,, if $\gamma \ll \omega_{\mathrm{L}}$, it is possible to distinguish between longitudinal and transverse components of the spin-density matrix. With Eq.~\eqref{markov_approximation} the inequality $\gamma \ll \omega_{\text{L}}$ yields the condition
\begin{equation}
	\label{markov_condition}
	\omega_{\mathrm{L}} \gg \frac{1}{T_{1}} 
\end{equation}
for the Larmor frequencies and Bloch times, for which it is permissible to replace the Lorentzian by an energy conserving $\delta$ function
\begin{equation}
	\label{lorentz_to_delta}
	\frac{\hbar \gamma}{\left(\Delta E_{j j'}\right)^2 + \left(\hbar \gamma\right)^2} \overset{\gamma \to 0}{\longrightarrow} \pi \, \delta( \Delta E_{j j'}).
\end{equation}
This reduces the numerical effort very considerably, because it allows one to eliminate an integration from the scattering term, and the energy conserving $\delta$ function is therefore often used without explicitly checking its validity. 

The considerations  leading to the connection between Eqs.~\eqref{markov_condition} and~\eqref{lorentz_to_delta} are a microscopic version of an argument due to Pines and Slichter,~\cite{Slichter} according to which $T_{1}$ and $T_{2}$ can differ only for correlation times that long in comparison to a Larmor period. The microscopic Boltzmann scattering terms, which contain the energy conserving $\delta$ functions and will be used in the following, do not apply in a regime outside of condition \eqref{markov_condition}. If $\omega_{\mathrm{L}} \simeq 1/T_{1}$, a finite broadening $\gamma$ has to be taken into account. Together the full equation of motion in the regime \eqref{markov_condition} for the itinerant-carrier spin density matrix thus reads 
\begin{widetext}
\begin{equation}
	\label{ph_Boltzmann}
	\begin{aligned}
		\frac{\partial}{\partial t} & \rho_{\vec{k}}^{\mu \mu'} = \frac{i}{\hbar} \big( \epsilon_{k}^{\mu} - \epsilon_{k}^{\mu'} \big)  \rho_{\vec{k}}^{\mu \mu'} \\
		&+ \frac{\pi}{\hbar} \sum_{k'} \sum_{\mu_{1} \mu_{2} \mu_{3}} M_{\vec{k}' \mu_{1}}^{\vec{k} \mu} M_{\vec{k} \mu_{3}}^{\vec{k}' \mu_{2}} \ \delta\big( \Delta E_{\vec{k}' \mu_2 \vec{k} \mu_3} \big) \left[ \big(1 + N^{\mathrm{ph}}_{|\vec{k}'-\vec{k}|}\big) \rho_{\vec{k}'}^{\mu_{1} \mu_{2}} \big(\delta_{\mu_{3} \mu'} - \rho_{\vec{k}}^{\mu_{3} \mu'}\big) - N^{\mathrm{ph}}_{|\vec{k}'-\vec{k}|} \rho_{\vec{k}}^{\mu_{3} \mu'} \big(\delta_{\mu_{1} \mu_{2}} - \rho_{\vec{k}'}^{\mu_{1} \mu_{2}}\big) \right] \\
		&- \frac{\pi}{\hbar} \sum_{\vec{k}'} \sum_{\mu_{1} \mu_{2} \mu_{3}} M_{\vec{k}' \mu_{1}}^{\vec{k} \mu} M_{\vec{k} \mu_{3}}^{\vec{k}' \mu_{2}} \ \delta\big( \Delta E_{\vec{k} \mu_3 \vec{k}' \mu_2} \big) \left[ \big(1 + N^{\mathrm{ph}}_{|\vec{k}-\vec{k}'|}\big) \rho_{\vec{k}}^{\mu_{3} \mu'} \big(\delta_{\mu_{1} \mu_{2}} - \rho_{\vec{k}'}^{\mu_{1} \mu_{2}}\big) - N^{\mathrm{ph}}_{|\vec{k}-\vec{k}'|} \rho_{\vec{k}'}^{\mu_{1} \mu_{2}} \big(\delta_{\mu_{3} \mu'} - \rho_{\vec{k}}^{\mu_{3} \mu'}\big) \right] \\
		&+ \frac{\pi}{\hbar} \sum_{\vec{k}'} \sum_{\mu_{1} \mu_{2} \mu_{3}} M_{\vec{k} \mu'}^{\vec{k}' \mu_{1}} M_{\vec{k}' \mu_{2}}^{\vec{k} \mu_{3}} \ \delta\big( \Delta E_{\vec{k}' \mu_2 \vec{k} \mu_3} \big) \left[ \big(1 + N^{\mathrm{ph}}_{|\vec{k}'-\vec{k}|}\big) \rho_{\vec{k}'}^{\mu_{2} \mu_{1}} \big(\delta_{\mu \mu_{3}} - \rho_{\vec{k}}^{\mu \mu_{3}}\big) - N^{\mathrm{ph}}_{|\vec{k}'-\vec{k}|} \rho_{\vec{k}}^{\mu \mu_{3}} \big(\delta_{\mu_{2} \mu_{1}} - \rho_{\vec{k}'}^{\mu_{2} \mu_{1}}\big) \right] \\
		&- \frac{\pi}{\hbar} \sum_{\vec{k}'} \sum_{\mu_{1} \mu_{2} \mu_{3}} M_{\vec{k} \mu'}^{\vec{k}' \mu_{1}} M_{\vec{k}' \mu_{2}}^{k \mu_{3}} \ \delta\big( \Delta E_{\vec{k} \mu_3 \vec{k}' \mu_2} \big) \left[ \big(1 + N^{\mathrm{ph}}_{|\vec{k}-\vec{k}'|}\big) \rho_{\vec{k}}^{\mu \mu_{3}} \big(\delta_{\mu_{2} \mu_{1}} - \rho_{\vec{k}'}^{\mu_{2} \mu_{1}}\big) - N^{\mathrm{ph}}_{|\vec{k}-\vec{k}'|} \rho_{\vec{k}'}^{\mu_{2} \mu_{1}} \big(\delta_{\mu \mu_{3}} - \rho_{\vec{k}}^{\mu \mu_{3}}\big) \right]
	\end{aligned}
\end{equation}
\end{widetext}
Here, $\Delta E_{\vec{k} \mu \vec{k}' \mu'} = \epsilon_{k}^{\mu} - \epsilon_{k'}^{\mu'} -  \hbar \omega^{\mathrm{ph}}_{|\vec{k}-\vec{k}'|}$, and $N^{\mathrm{ph}}_{q}$ is the occupation function of a thermalized phonon bath, given by a Bose-Einstein distribution
\begin{equation}
	\label{phDist_eq}
		 N^{\mathrm{ph}}_{q} =  \frac{1}{ e^{\beta \hbar \omega_{\mathrm{ph}}}-1 }
\end{equation}
where $\beta = 1/(k_{\mathrm{B}} T_{\mathrm{ph}})$.
The numerical results for the microscopic dynamics in the following sections are obtained by numerically solving the equations of motion \eqref{locEoM_Kondo} and \eqref{ph_Boltzmann}. In the numerical calculations the spin-density matrix is transformed to the single-particle basis of the instantaneous, diagonalized eigenstates~\eqref{itin_eigenstates}.

\section{Dephasing between localized and itinerant spins: numerical results \label{simulation_iso}}

Since we are interested in this paper in a comparison of the model described above with Landau-Lifshitz and Gilbert damping, we investigate magnetization dynamics with an initial spin-density matrix that corresponds to a tilting of the spins out of their equilibrium position \emph{without} changing the kinetic energy of the carriers, because we need initial conditions that lead to generic magnetization dephasing without carrier heating and the corresponding demagnetization dynamics.

\subsection{Initial state and spin dynamics \label{initial-state}}

\begin{figure}[tb]
	\centering
		\includegraphics[width=0.45\textwidth]{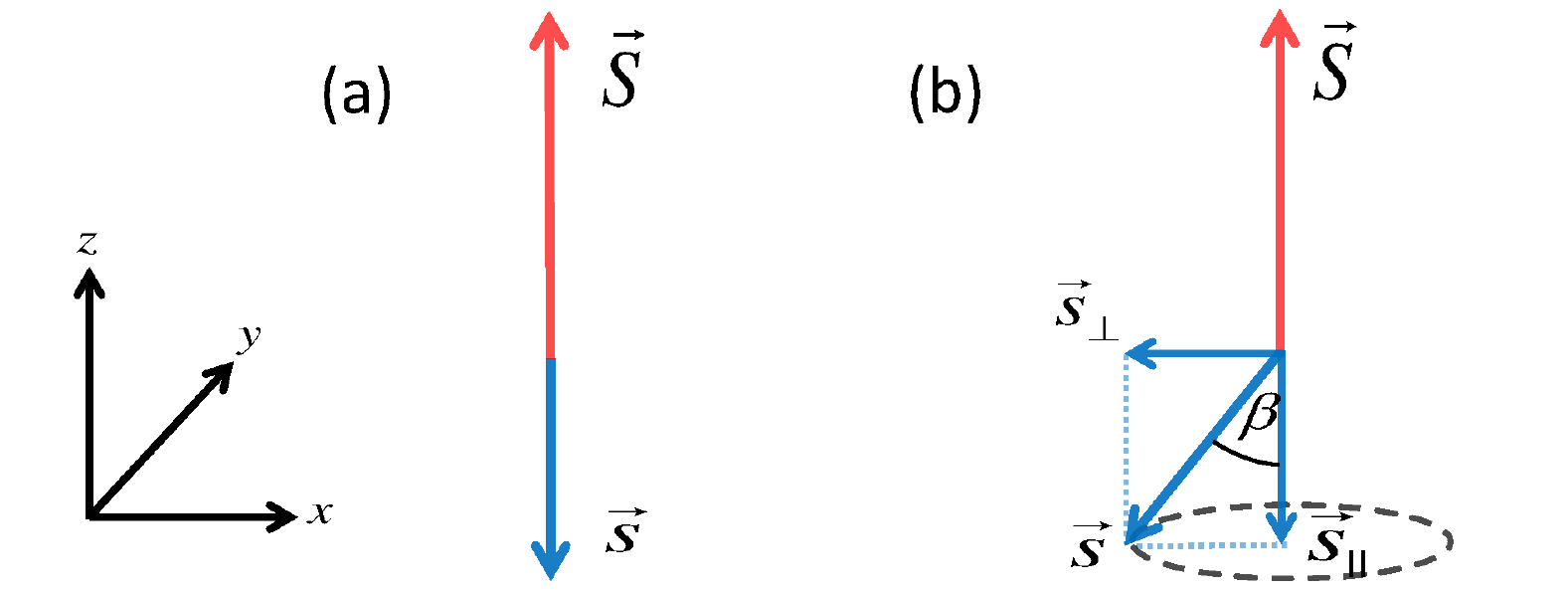}
	\caption{ (a) Localized spin $\langle \vec S\rangle $ and itinerant spin $\langle \vec s\rangle $ in thermal equilibrium. (b) Itinerant spins tilted out of equilibrium by an angle $\beta$.}
	\label{fig:initial_state}
\end{figure}

Thus we take as the equilibrium initial state the steady-state that is reached for the coupled spins interacting with the phonon bath at low temperature ($T_{\mathrm{ph}}=1 \, \mathrm{K}$), as shown in Fig.~\ref{fig:initial_state}(a). In this equilibrium state the spin density matrix is characterized by shifted Fermi functions for the distributions $\rho^{\mu\mu}_{\vec{k}}=f(\epsilon^{\mu}_{k}-E_{\mathrm{F}}, T_{\mathrm{ph}})$ and vanishing coherences $\rho^{+-}_{\vec{k}}=0$. In particular, the steady-state calculation determines the equilibrium magnetic gap.

\begin{figure}[tb]
	\centering
		\includegraphics[width=0.49\textwidth]{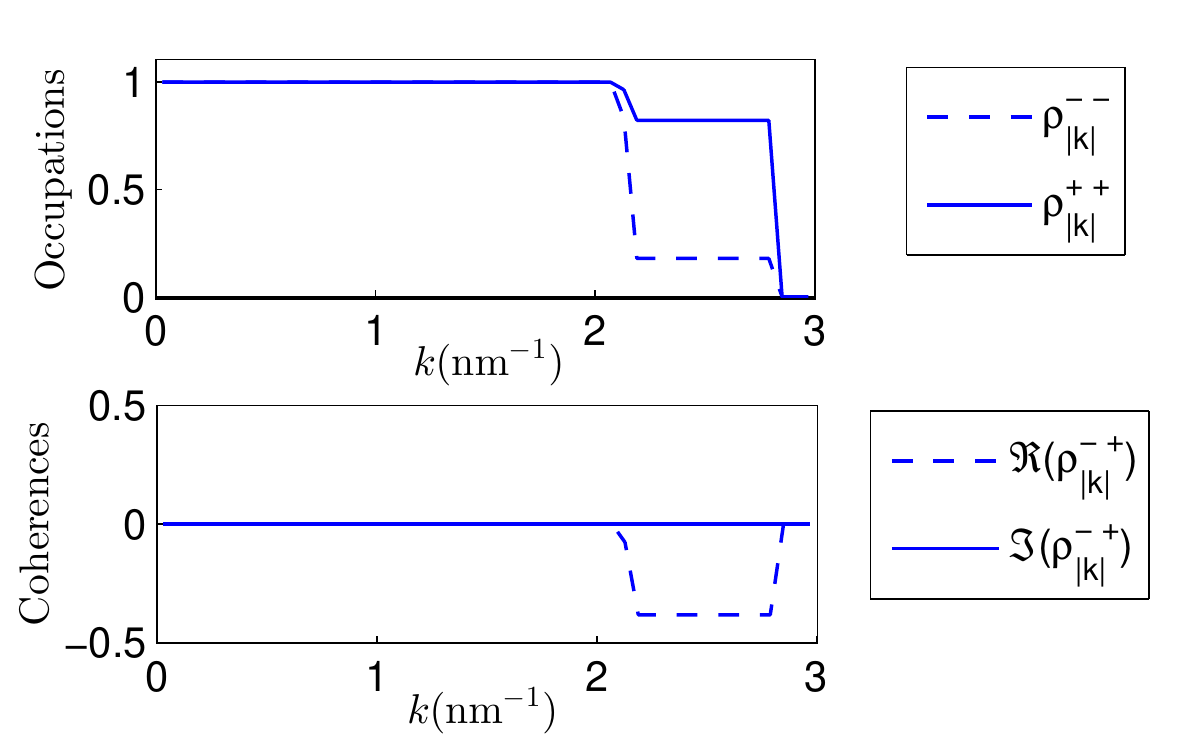}
	\caption{Initial spin density-matrix (occupations and coherences) for itinerant electrons corresponding to a tilt~$\beta = 50^{\circ}$. The deviation from equilibrium occurs only between the Fermi wave-vectors of the ``$+$'' and ``$-$'' bands.}
	\label{fig:distributions}
\end{figure}

We then change the itinerant density matrix to that corresponding to an itinerant spin tilted by $\beta = 10^{\circ}$ out of equilibrium, see Fig.~\ref{fig:initial_state}(b). This initial condition achieves a tilting of the spins without heating and avoids generic de- and remagnetization dynamics. Microscopically the tilted spin corresponds to the spin density-matrix shown in Fig.~\ref{fig:distributions}. The perturbation for distributions and coherences exists only between the two Fermi wave vectors for the $\mu=+$ and $\mu=-$ bands. For smaller tilt angles the deviation is much less pronounced.


\begin{figure}[tb]
	\centering
		\includegraphics[width=0.45\textwidth]{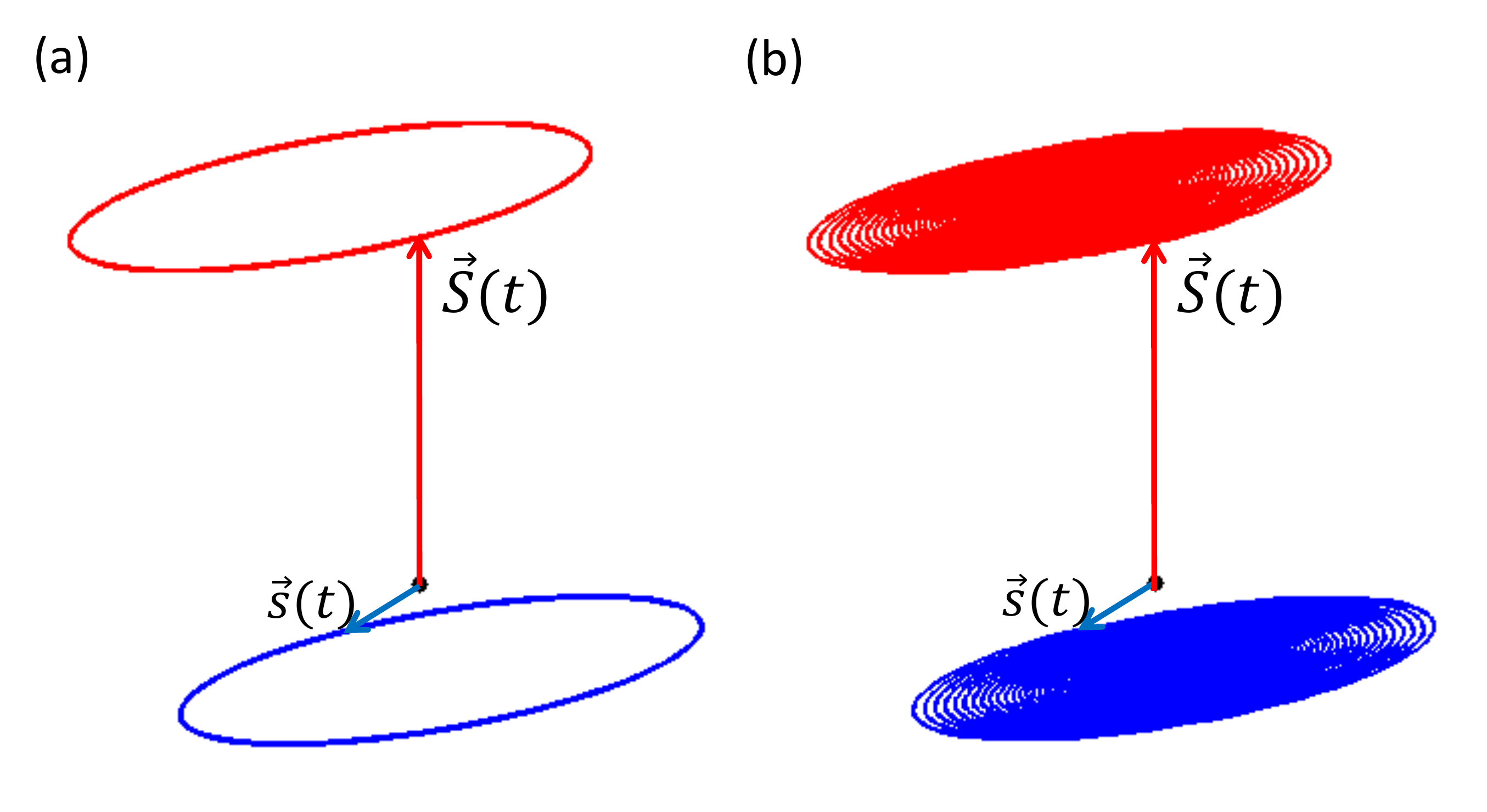}
	\caption{Non-equilibrium dynamics of localized (red) and itinerant (blue) spins including electron-phonon scattering without spin-mixing (a) and within spin-mixing (b).}
	\label{fig:so_3D}
\end{figure}

From this initial condition, both spins start to precess around the instantaneous direction, along which the exchange interaction tries to align them. This direction is determined for the itinerant carriers by the localized spins, and vice versa. A return back into equilibrium requires the scattering of itinerant electrons with phonons. If we switch off spin mixing, the dynamics shown in Fig.~\ref{fig:so_3D}(a) result: No angular momentum is exchanged with the phonon bath, the excited system cannot relax into equilibrium and the precession goes on indefinitely. Fig.~\ref{fig:so_3D}(b) shows the same result including spin-mixed itinerant states. Now angular momentum can be transferred from the itinerant sub-system into the lattice and the total spin $\langle \vec{S}\rangle +\langle \vec{s}\rangle$ changes. In the presence of spin-orbit coupling, electron-phonon scattering, which is by itself spin-diagonal, can return the spin system into equilibrium, characterized by aligned spins and vanishing transverse components. Since we consider here a small Rashba coupling, the interaction with the phonon bath removes energy much faster than angular momentum. The carrier temperature therefore stays practically equal to the phonon temperature $T_{\mathrm{ph}}$ during the entire relaxation process, and no heat-induced demagnetization processes occur. The final magnetization is, however, not necessarily oriented in the $z$ direction of the fixed frame, because in the results discussed in the this section there is no external field or anisotropy to induce such an alignment.

\begin{figure}[tb]
	\centering
		\includegraphics[width=0.47\textwidth]{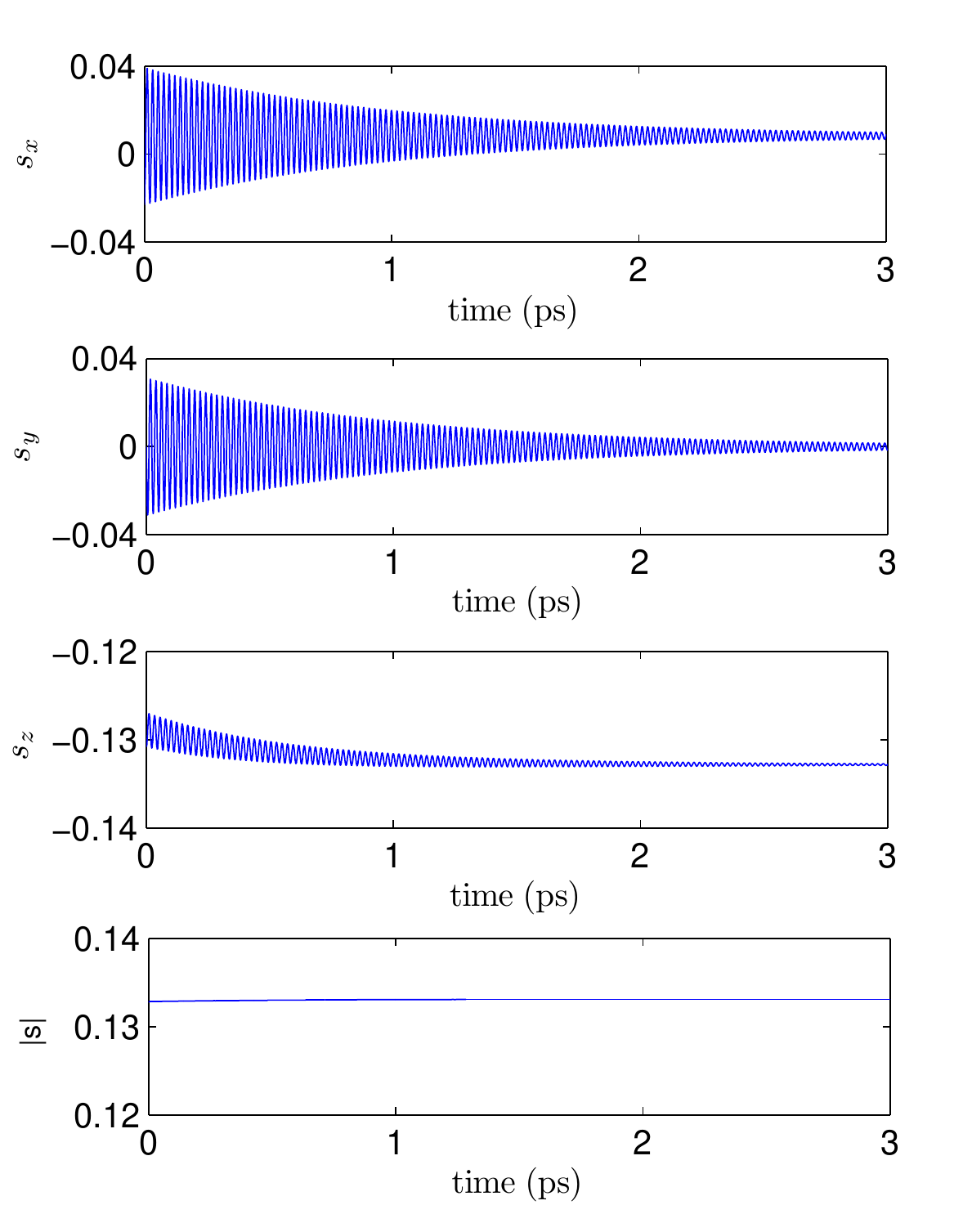}
	\caption{Relaxation dynamics of itinerant spins in the fixed frame.}
	\label{fig:fixed_frame}
\end{figure}

\begin{figure}[tb]
	\centering
		\includegraphics[width=0.47\textwidth]{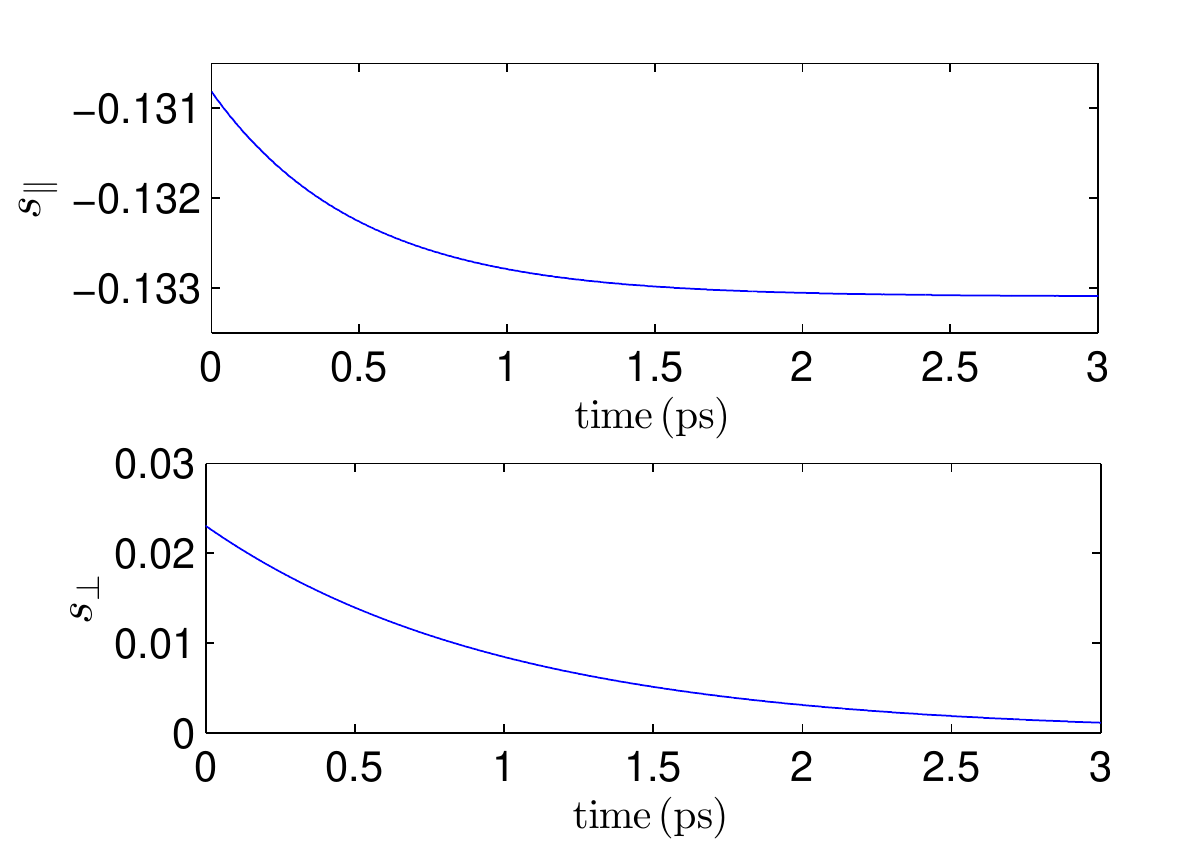}
	\caption{Relaxation dynamics of itinerant spins in the rotating frame. An exponential fit determines the Bloch decay times to $T_1 = 0.5 \, \mathrm{ps}$ and $T_2 = 1.0 \, \mathrm{ps}$.}
	\label{fig:rot_frame}
\end{figure}

We plot the resulting dynamics of the itinerant spins during the dephasing process in Fig.~\ref{fig:fixed_frame}, which shows that all itinerant spatial components precess, and no spatially fixed component can be considered to be longitudinal. First, the localized spin turns away from the $z$ direction due to the tilted itinerant spin and subsequently both spin-systems precess around each other because the quantization axis of each system changes continuously due to the mutual interaction. During the entire relaxation process the absolute value of the itinerant spin is conserved to better than 1\%. Fig.~\ref{fig:rot_frame} shows the same dynamics in the rotating frame, where the longitudinal and transverse dynamics can be seen clearly.
Both itinerant components show exponential dynamics, which are therefore well described by decay times $T_{1}$ and $T_{2}$. If well-defined decay times exist, one expects a ratio $T_{2}/T_{1}=2$ as long as the length of the spin is conserved. The fit for our numerical results indeed gives $2 T_{1} \simeq T_{2}$. 

\begin{figure}[tb]
	\centering
		\includegraphics[width=0.47\textwidth]{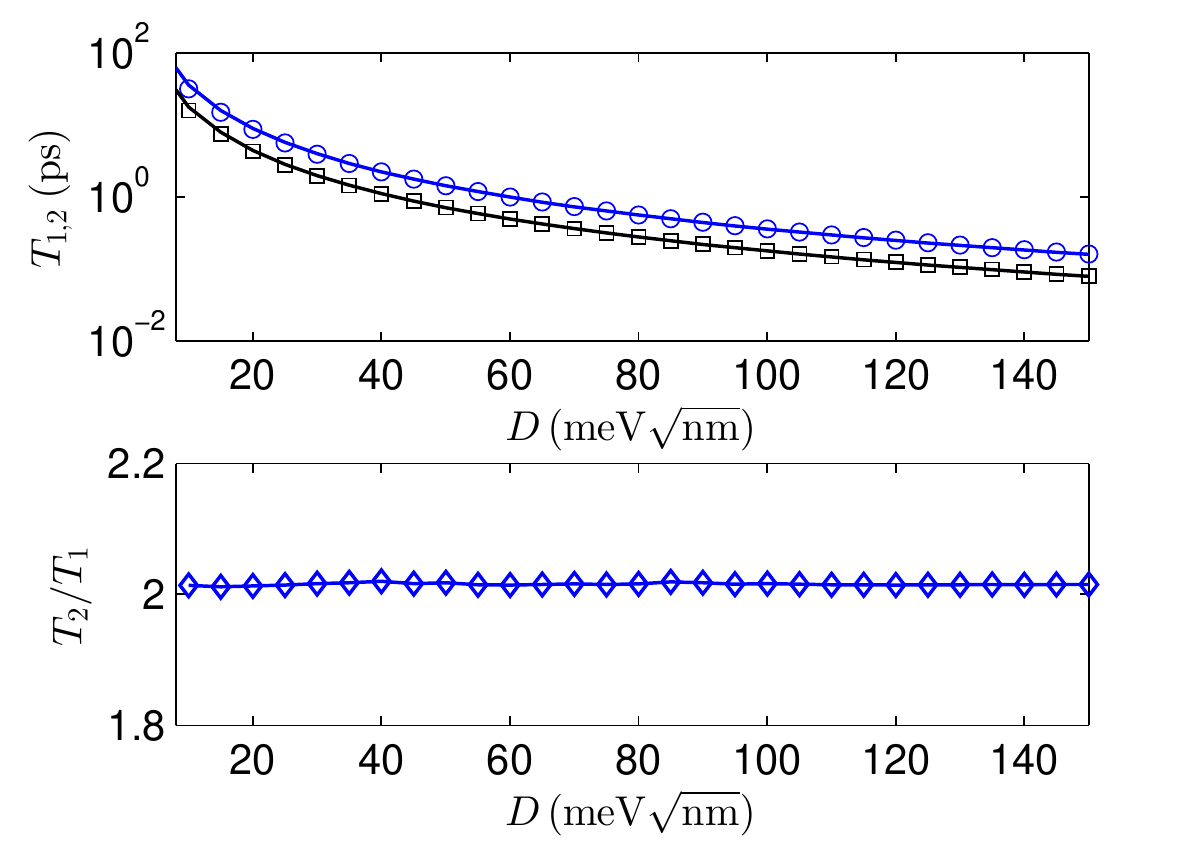}
	\caption{Top: Longitudinal (black squares) and the transverse (blue circles) Bloch decay times vs.\  electron-phonon coupling strength $D$. Two fit curves $\propto 1/D^2$ show that $T_{1,2} \propto 1/D^2$. Bottom:The ratio of both decay times remains almost constant $T_{2}/T_{1} \approx 2$ over the entire range of coupling. }
	\label{fig:M0_T12}
\end{figure}

Figure~\ref{fig:M0_T12} plots the $T_1$ and $T_2$ values extracted from the dynamics as a function of the strength of the electron-phonon coupling, or deformation potential, $D$. The dependence of the decay times on $D$ can be fit extremely well by a $1/D^2$ relation, which demonstrates the proportionality of the Bloch decay times $T_{1,2} \propto 1/D^{2}$. Further, the ratio $T_2/T_1$ stays equal to 2 for all coupling strengths. As discussed in Sec.~\ref{eph_scattering} about the Markov approximation, our microscopic description cannot reach regimes where $T_{1}$ and $T_{2}$ are indistinguishable and therefore equal. However, these results show that for small tilting angles, not even a pronounced electron-phonon coupling leads to a noticeable deviation from the $T_2=2T_1$ behavior. Because this relation between $T_1$ and $T_2$ holds, the dynamics in Fig.~\ref{fig:rot_frame} can be equally well described by an Landau-Lifshitz or Gilbert damping term. By fitting the dephasing time $T_{2} \approx 1 \, \mathrm{ps}$ and the Larmor-frequency $\omega_{\mathrm{L}} \approx 281 \, \mathrm{ps^{-1}}$, equation \eqref{alpha_T2_relation} yields the corresponding Gilbert damping parameter $\alpha_{\mathrm{iso}} \approx 3.6 \cdot 10^{-3}$. 

\subsection{Precession-frequency shift due to dephasing \label{varying_larmor}}

\begin{figure}[tb]
	\centering
		\includegraphics[width=0.48\textwidth]{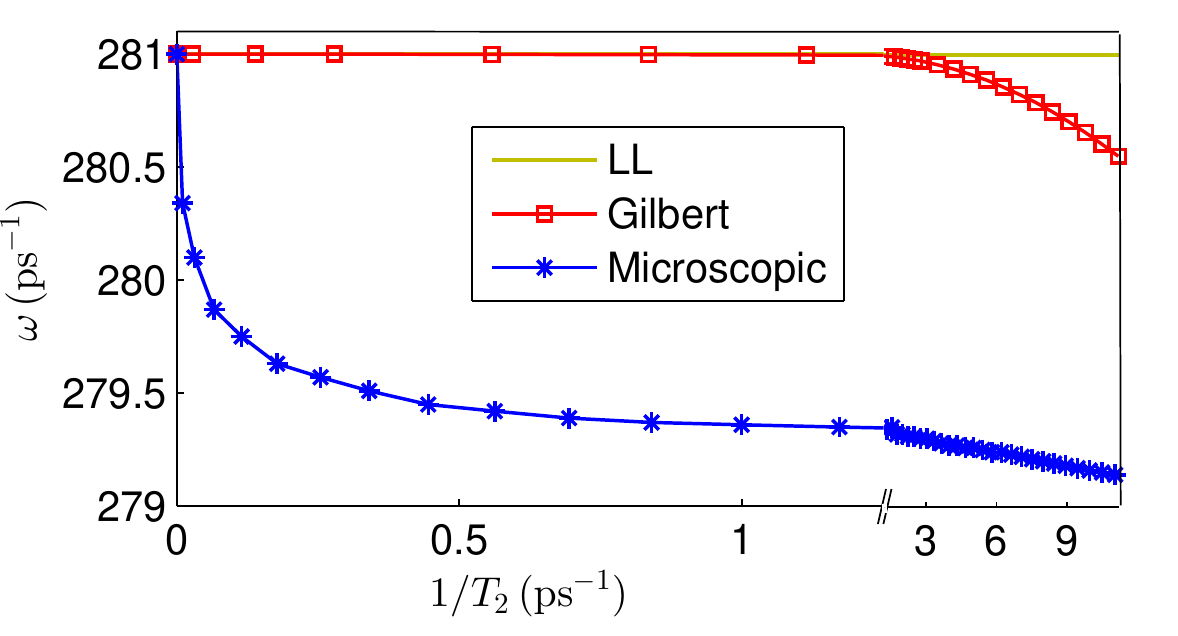}
	\caption{Precession frequency with respect to the damping strength in terms of the decay rate $1/T_{2}$.}
	\label{fig:larmor}
\end{figure}

In the Landau-Lifshitz equation the contributions describing, respectively, the precession and the damping are completely independent, so that the damping constant $\lambda$ has no impact on the precession. By contrast, an increase of $\alpha$ in the Gilbert equation does not only increase the dephasing rate, it lowers the precession frequency as well. 

In this section we investigate the change of the precession frequency in the microscopic calculation and compare it with the macroscopic descriptions. To this end, we use the off-diagonal components of the itinerant density matrix $\rho^{+-}(t) = \sum_{\vec{k}} \rho_{\vec{k}}^{+-}(t)$, which describe the dynamics of the transverse spin components in the rotating frame. The modulus of its Fourier transform $|\rho^{+-}(\omega)|$ shows a distinct peak, which is exactly at the precession frequency.

To compare the precession frequency for different damping parameters, we use the dependence on $T_{2}$, because all dephasing parameters can be related to $T_2$ for small excitations. Fig.~\ref{fig:larmor} plots the Larmor frequency vs. the transverse relaxation rate $1/T_{2}$. The precession parameters of the Landau-Lifshitz and the Gilbert equation are chosen such that the Larmor frequency in the undamped limiting case is equal to that of the microscopic simulation. In order to stay within the bounds set by condition~\eqref{markov_condition}, we do not extend the plot in Fig.~\ref{fig:larmor} to higher dephasing rates. Fig.~\ref{fig:larmor} shows that the microscopic calculation yields a reduction of the precession frequency with the damping rate. Although the Gilbert dynamics also show such a reduction, it occurs only at shorter $T_2$. As mentioned above, for the Landau-Lifshitz damping, the frequency is independent of the damping parameter. Even though the change of precession frequency in the microscopic calculation is small, the Landau-Lifshitz damping completely fails to include this effect. While Gilbert damping does show a reduction of precession frequency, it is not at all close to the microscopic calculation on the frequency scale considered here. Both phenomenological damping expressions thus do not reproduce the dependence of the precession frequency on $T_2$. Even though the numerical differences are small, these differences already occur in the small-excitation regime, and may perhaps be detectable.


\subsection{Dephasing at larger excitation angles \label{varying_angle}}

The phenomenological Landau-Lifshitz and Gilbert damping contributions describe an exponential decay only for small excitation angles, as studied in the previous section.
In this section we investigate the effect of larger excitation angles ($>10^{\circ}$) on the spin dynamics in the microscopic calculation. Apart from this the initial condition of the dynamics is the same as before, in particular, the itinerant spin is tilted such that the absolute value of the spin is unchanged.

\begin{figure}[tb]
	\centering
	\includegraphics[width=0.49\textwidth]{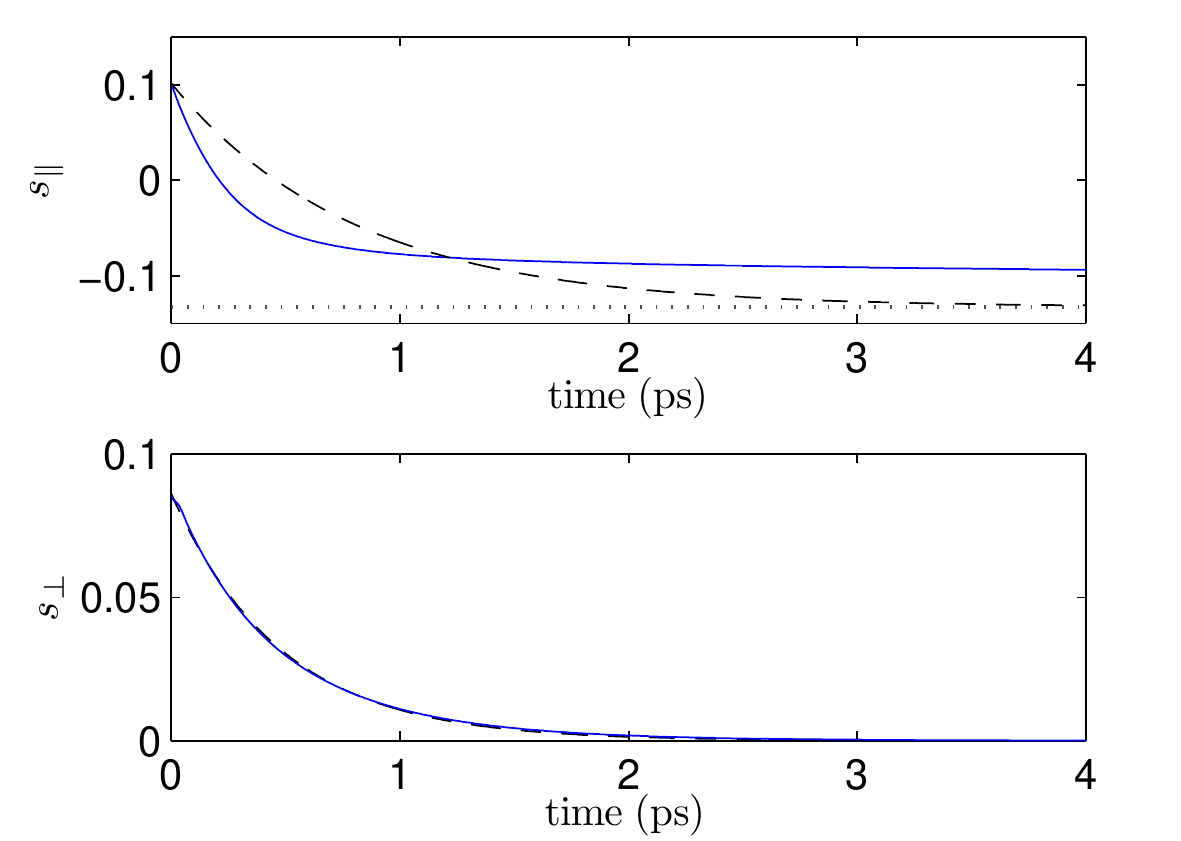}
	\caption{Dynamics of the longitudinal and transverse itinerant spin components in the rotating frame (solid lines) for a tilt angle of $\beta=140^{\circ}$, together with exponential fits toward equilibrium (dashed lines). The longitudinal equilibrium polarization is shown as a dotted line.}
	\label{fig:rotframe_140}
\end{figure}

Figure~\ref{fig:rotframe_140} shows the time development of the  $s_\|$ and $s_\perp$ components of the itinerant spin in the rotating frame for an initial tilt angle $\beta = 140^\circ$.  While the transverse component $s_\perp$ in the rotating frame can be well described by an exponential decay,  the longitudinal component $s_\|$ shows a different behavior. It initially decreases with a time constant of less than $1 \, \mathrm{ps}$, but does not reach its equilibrium value. Instead, the eventual return to equilibrium takes place on a much longer timescale, during which the $s_\perp$ component is already vanishingly small. The long-time dynamics are therefore purely collinear. For the short-time dynamics, the transverse component can be fit well by an exponential decay, even for large excitation angles. This behavior is different from Landau-Lifshitz and Gilbert dynamics, cf.~Fig.~\ref{fig:LL_fig}, which both exhibit non-exponential decay of the transverse spin component.

\begin{figure}[tb]
	\centering
	\includegraphics[width=0.48\textwidth]{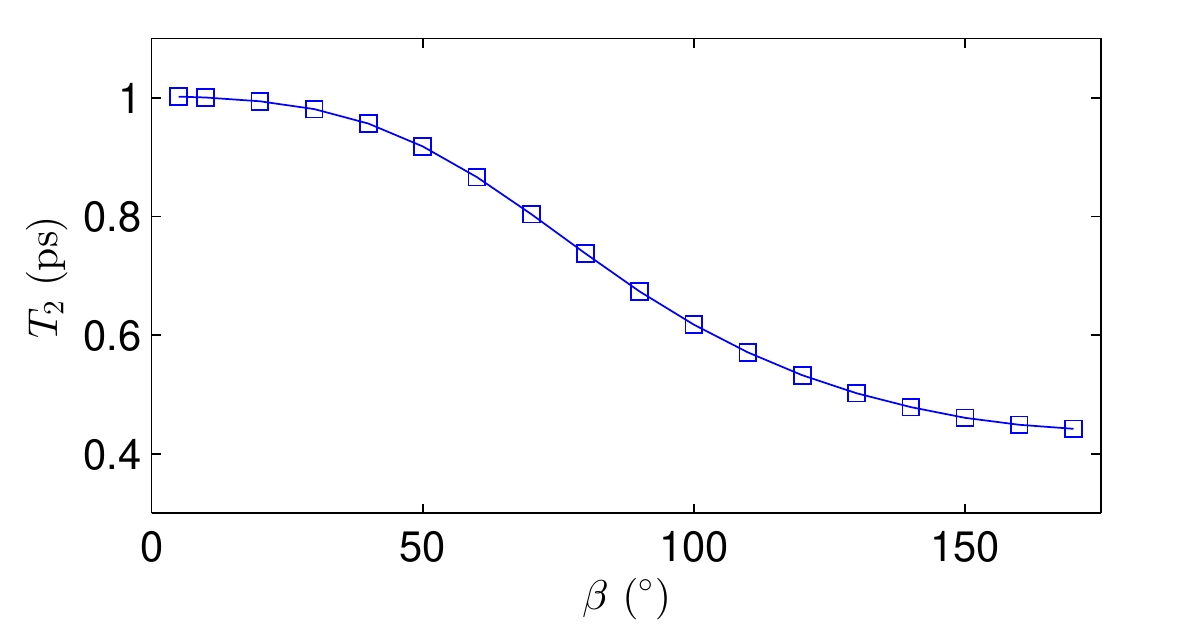}
	\caption{$T_2$ time extracted from exponential fit to $s_\perp$ dynamics in rotating frame for different initial tilting angles~$\beta$.}
	\label{fig:T12_vs_angle}
\end{figure}

In Fig.~\ref{fig:T12_vs_angle} the dependence of $T_2$ on the excitation angle is shown. From small $\beta$ up to almost $180^\circ$, the decay time decreases by more than 50\%. This dependence is exclusively due to the ``excitation condition,'' which involves only spin degrees of freedom (``tilt angle''), but no change of temperature. Although one can fit such a $T_2$ time to the transverse decay, the overall behavior with its two stages is, in our view, qualitatively different from the typical Bloch relaxation/dephasing picture.  

\begin{figure}[tb]
	\centering
	\includegraphics[width=0.45\textwidth]{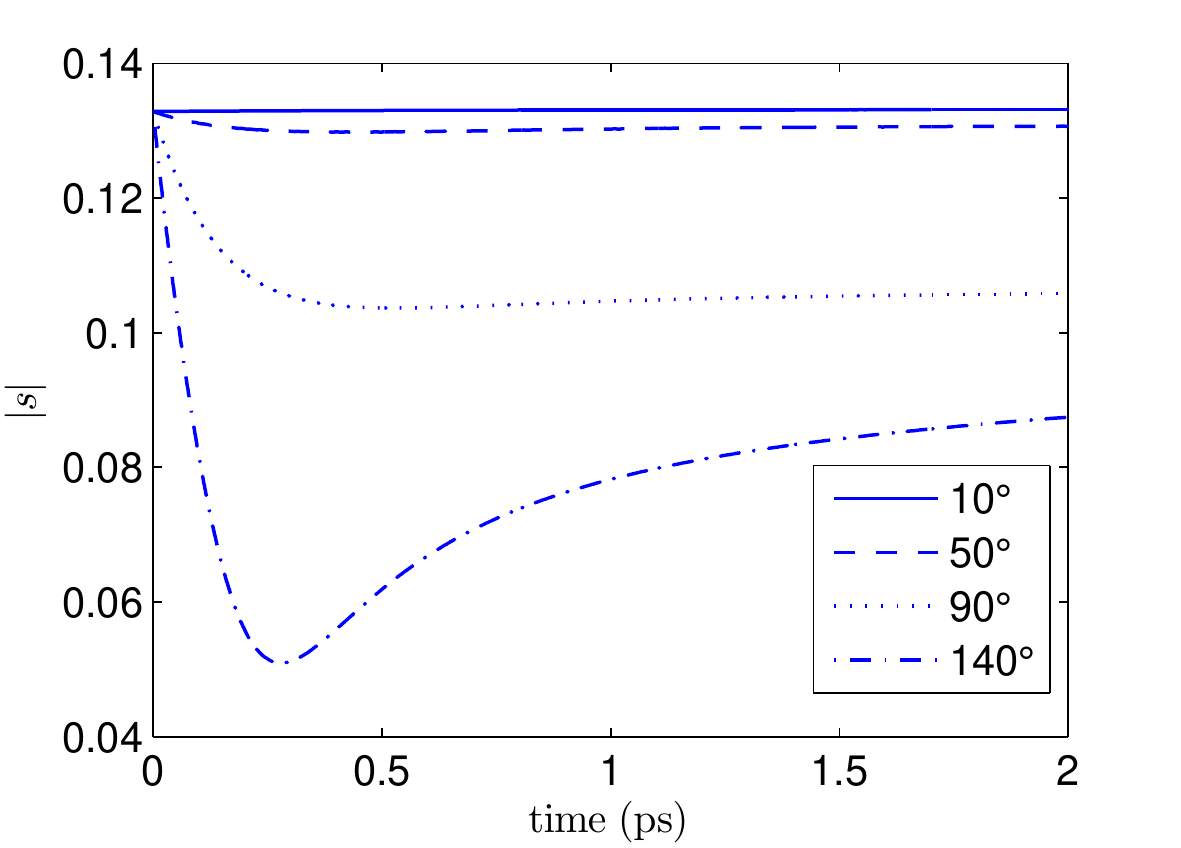}
	\caption{Dynamics of the modulus $|\vec s|$ of the itinerant spin for different initial tilt angles $\beta$. Note the slightly different time scale compared to Fig.~\ref{fig:rotframe_140}.}
	\label{fig:s_abs_angle}
\end{figure}

To highlight the similarities and differences from the Bloch relaxation/dephasing we plot in Fig.~\ref{fig:s_abs_angle} the modulus of the itinerant spin vector $|\vec{s}|$ in the rotating frame, whose transverse and longitudinal components were shown in Fig.~\ref{fig:rotframe_140}. Over the 2\,ps, during which the transverse spin in the rotating frame essentially decays, the modulus of the spin vector undergoes a fast initial decrease and a partial recovery. The initial length of $\vec s$ is recovered only over a much larger time scale of several hundred picoseconds (not shown). Thus the dynamics can be seen to differ from a Landau-Lifshitz or Gilbert-like scenario because the spin does not precess toward equilibrium with a constant length. Additionally they differ from Bloch-like dynamics because there is a combination of the fast and slow dynamics that cannot be described by a single set of $T_1$ and $T_2$ times. We stress that the microscopic dynamics at larger excitation angles show a precessional motion of the magnetization without heating and a slow remagnetization. This scenario is somewhat in between typical small angle-relaxation, for which the modulus of the magnetization is constant and which is well described by Gilbert and Landau-Lifshitz damping, and collinear de/remagnetization dynamics. 

\section{Effect of anisotropy\label{simulation_aniso}}

So far we have been concerned with the question how phenomenological equations describe dephasing processes between itinerant and localized spins, where the magnetic properties of the system were determined by a mean-field exchange interaction only. Oftentimes, phenomenological models of spin dynamics are used to describe dephasing processes toward an ``easy axis'' determined by anisotropy fields.~\cite{Vonsovskii}

In order to capture in a simple fashion the effects of anisotropy on the spin dynamics in our model, we simply assume the existence of an effective anisotropy field $\vec{H}_{\mathrm{aniso}}$, which enters the Hamiltonian via 
\begin{equation}
	\label{aniso_Ham}
		 \hat{\mathcal{H}}_{\mathrm{aniso}} = - g \mu_{\mathrm{B}} \mu \; \hat{\vec{s}} \cdot \vec{H}_{\mathrm{aniso}} 
\end{equation}
and only acts on the itinerant carriers. Its strength is assumed to be small in comparison to the field of the localized moments $\vec{H}_{\mathrm{loc}}$. This additional field $\vec{H}_{\mathrm{aniso}}$ has to be taken into account in the diagonalization of the coherent dynamics as well, see section~\ref{Basis}. 

\begin{figure}[tb]
	\centering
		\includegraphics[width=0.45\textwidth]{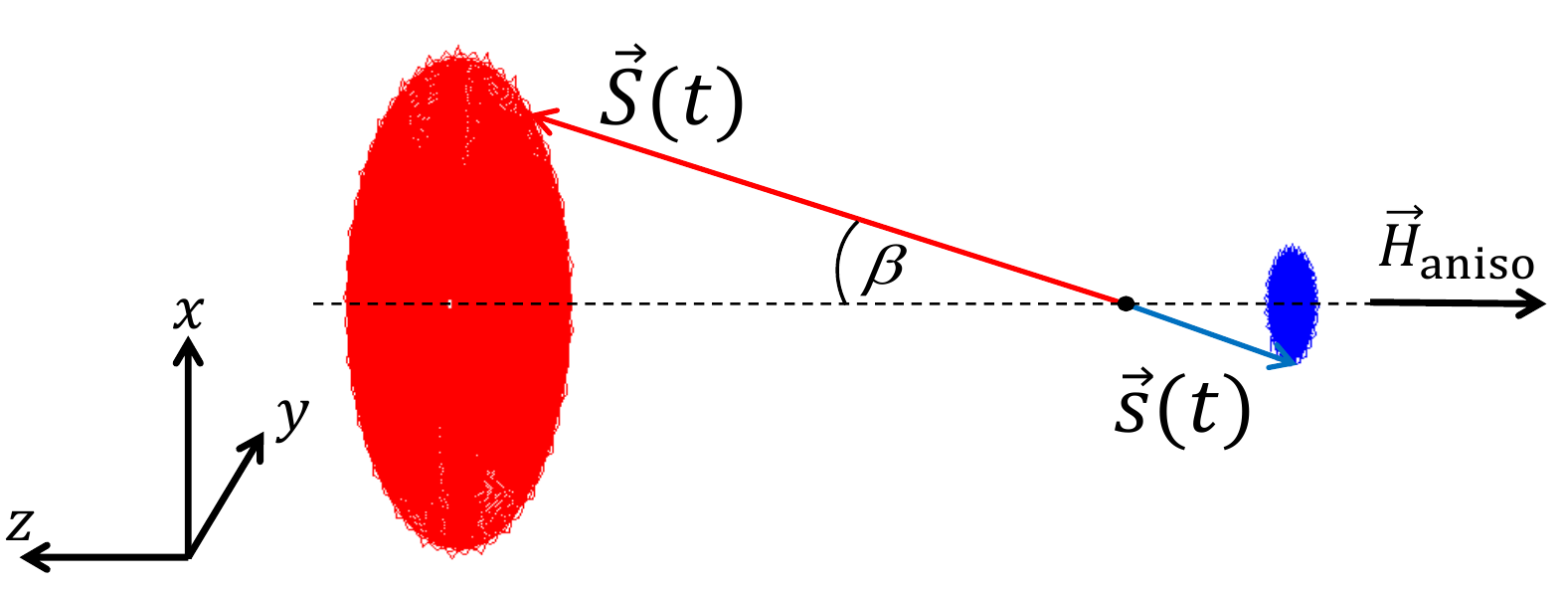}
	\caption{Dynamics of the localized spin $\vec{S}$ and itinerant spin $\vec{s}$. At $t=0$, the equilibrium configuration of both spins is tilted ($\beta = 10^{\circ}$) with respect to an anisotropy field $\vec{H}_{\mathrm{aniso}}$. The anisotropy field is only experienced by the itinerant sub-system.}
	\label{fig:aniso_3D}
\end{figure}

For the investigation of the dynamics with anisotropy, we choose a slightly different initial condition, which is shown in Fig.~\ref{fig:aniso_3D}. In thermal equilibrium, both spins are now aligned, with opposite directions, along the anisotropy field $\vec{H}_{\mathrm{aniso}}$, which is assumed to point in the $z$ direction. At $t=0$ they are both rigidly tilted by an angle $\beta = 10^\circ$ with respect to the anisotropy field. 

Figure~\ref{fig:aniso_3D} shows the time evolution of both spins in the fixed frame, with $z$ axis in the direction of the anisotropy field for the same material parameters as in the previous sections and an anisotropy field $\vec{H}_{\mathrm{aniso}} = - 10^{8} \, \frac{A}{m} \cdot \vec{e}_{z}$. The dynamics of the entire spin-system are somewhat different now, as the itinerant spin precesses around the combined field of the anisotropy and the localized moments. The localized spin precesses around the itinerant spin, whose direction keeps changing as well. 

\begin{figure}[tb]
	\centering
		\includegraphics[width=0.48\textwidth]{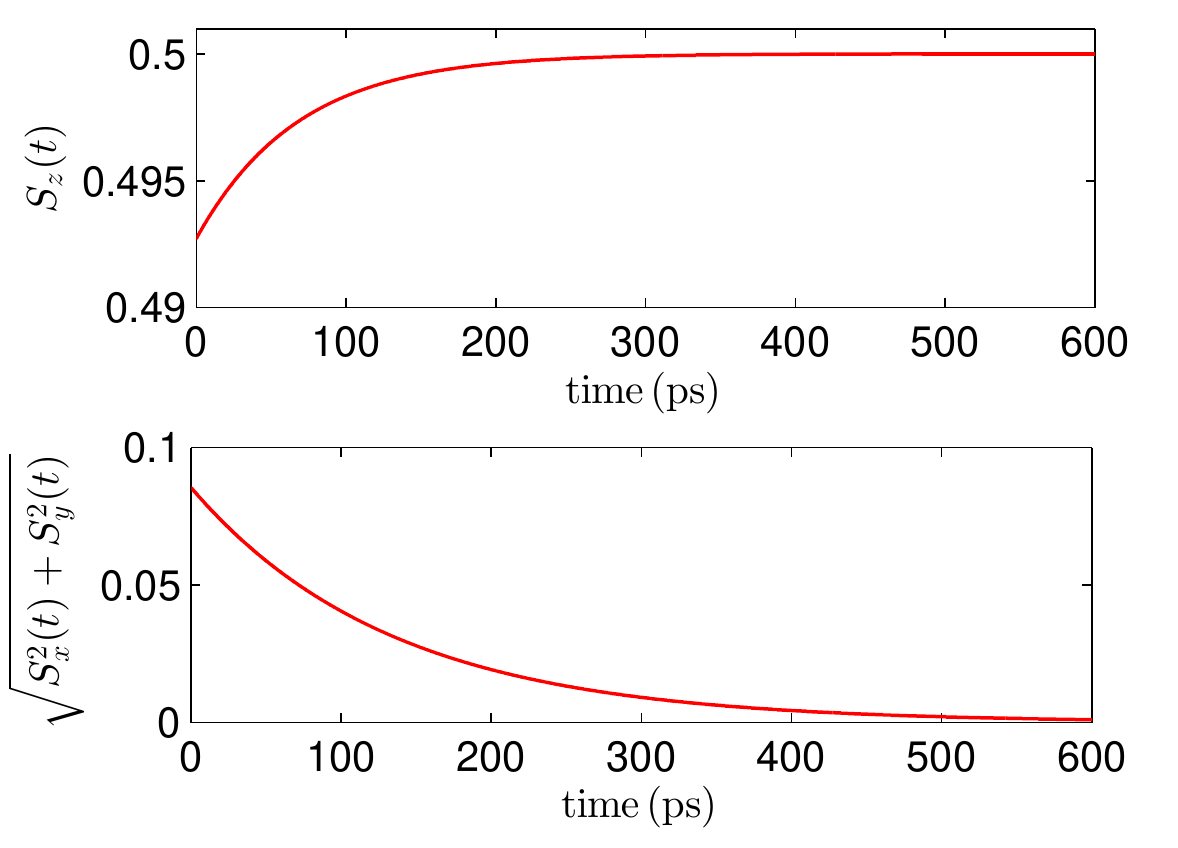}
	\caption{Relaxation dynamics of the localized spin toward the anisotropy direction for longitudinal component $S_{z}$ and the transverse component $\sqrt{S_x^2 +S_y^2}$. An exponential fit yields Bloch decay times of $T^{\mathrm{aniso}}_1 = 67.8 \, \mathrm{ps}$ and $T^{\mathrm{aniso}}_2 = 134.0 \, \mathrm{ps}$. }
	\label{fig:aniso_locdyn}
\end{figure}

Figure~\ref{fig:aniso_locdyn} contains the dynamics of the components of the localized spin in the rotating frame. Both components show an exponential behavior that allows us to extract well defined Bloch-times $T^{\mathrm{aniso}}_{1}$ and $T^{\mathrm{aniso}}_{2}$. Again we find the ratio of $2 T^{\mathrm{aniso}}_{1} \approx T^{\mathrm{aniso}}_{2}$, because the absolute value of the localized spin does not change, as it is not coupled to the phonon bath. 

\begin{figure}[tb]
	\centering
		\includegraphics[width=0.49\textwidth]{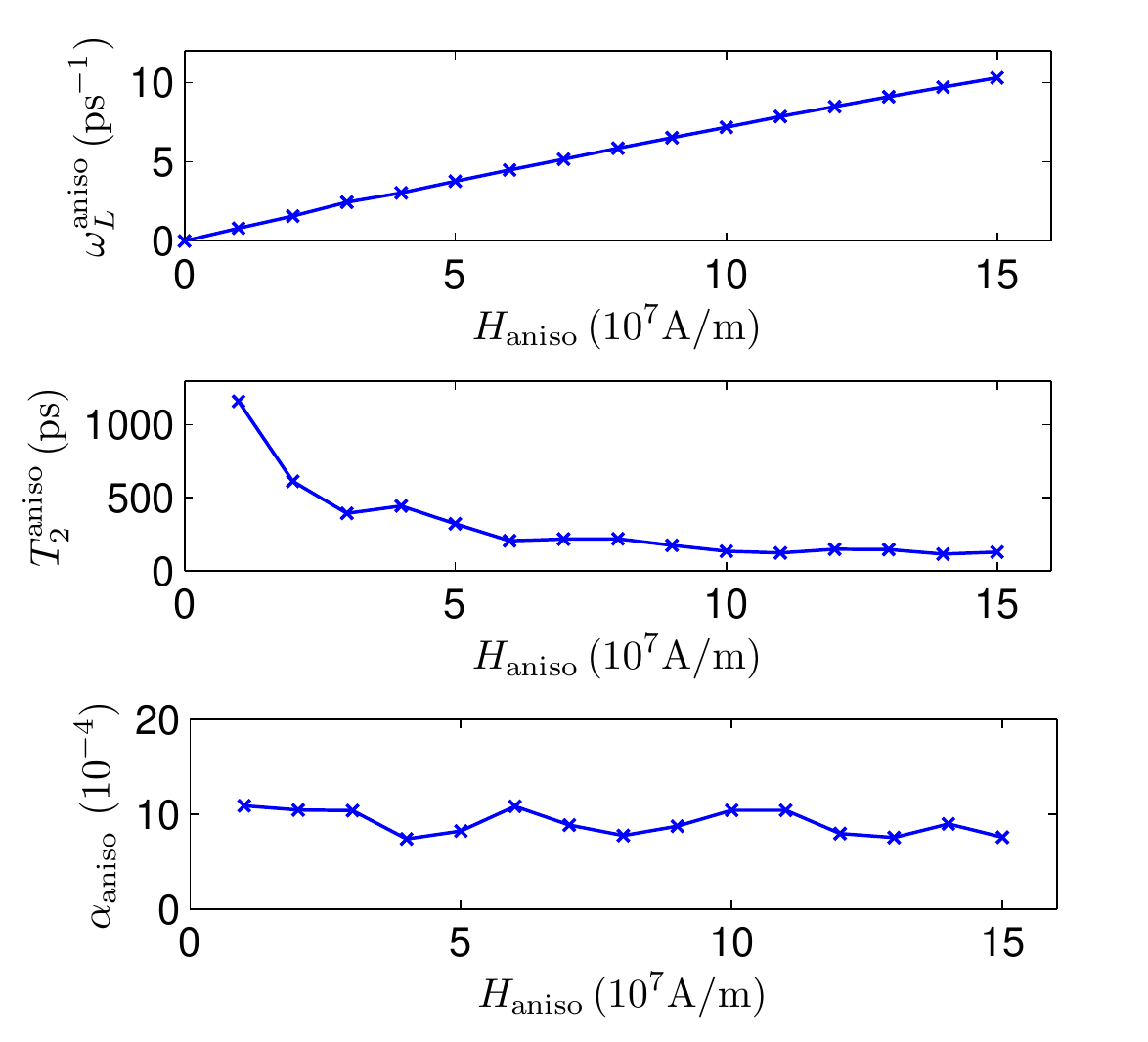}
	\caption{Larmor frequency $\omega^{\mathrm{aniso}}_{\mathrm{L}}$ and Bloch decay time $T^{\mathrm{aniso}}_{2}$ extracted from the spin dynamics vs.~anisotropy field $H_{\mathrm{aniso}}$, as well as the corresponding damping parameter $\alpha_{\mathrm{aniso}}$.}
	\label{fig:aniso_alpha_Hext}
\end{figure}

In Fig.~\ref{fig:aniso_alpha_Hext} the Larmor-frequency $\omega_{\mathrm{L}}^{\mathrm{aniso}}$, which is the precession frequency due to the anisotropy field, and the Bloch decay times $T^{\mathrm{aniso}}_{2}$ are plotted vs.~the strength of the anisotropy field $\vec{H}_{\mathrm{aniso}}$.  The Gilbert damping parameter $\alpha_{\mathrm{aniso}}$ for the dephasing dynamics computed via Eq.~\eqref{alpha_T2_relation} is also presented in this figure.

The plot reveals a decrease of the dephasing time $T^{\mathrm{aniso}}_{2}$ and a almost linear increase of the Larmor frequency $\omega^{\mathrm{aniso}}_{\mathrm{L}}$ with the strength of the anisotropy field $H_{\mathrm{aniso}}$. The Gilbert damping parameter $\alpha_{\mathrm{aniso}}$ shows only a negligible dependence on the anisotropy field $H_{\mathrm{aniso}}$. This confirms the statement that, in contrast to the dephasing rates, the Gilbert damping parameter is independent of the applied magnetic field. In the investigated range we find an almost constant value of $\alpha_{\mathrm{aniso}} \simeq 9 \times 10^{-4}$. 

The  Gilbert damping parameter $\alpha_{\mathrm{aniso}}$ for the dephasing toward the anisotropy field is about 4 times smaller than $\alpha_{\mathrm{iso}}$, which describes the dephasing between both spins. This disparity in the damping efficiency ($\alpha_{\mathrm{aniso}} < \alpha_{\mathrm{iso}}$) is obviously due to a fundamental difference in the dephasing mechanism. In the anisotropy case the localized spin dephases toward the $z$ direction without being involved in scattering processes with itinerant carriers or phonons. The dynamics of the localized spins is purely precessional due to the time-dependent magnetic moment of the itinerant carriers $\vec{H}_{\mathrm{itin}}(t)$. Thus, only this varying magnetic field, that turns out to be slightly tilted against the localized spins during the entire relaxation causes the dephasing, in presence of the coupling between itinerant carriers and a phonon bath, which acts as a sink for energy and angular momentum. The relaxation of the localized moments thus occurs only indirectly as a carrier-meditated relaxation via their coupling to the time dependent mean-field of the itinerant spin.

\begin{figure}[tb]
	\centering
		\includegraphics[width=0.49\textwidth]{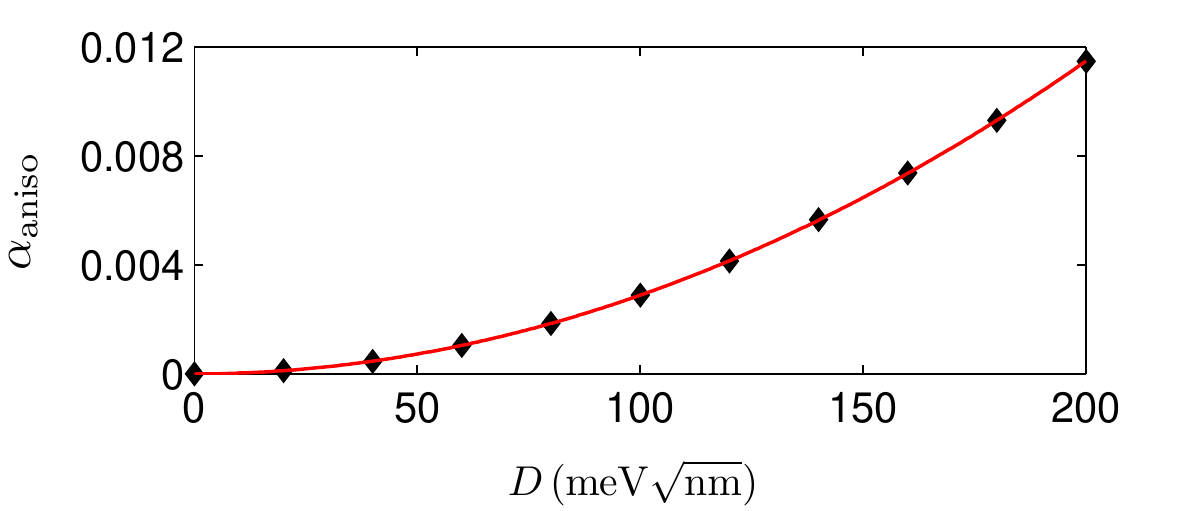}
	\caption{Damping parameter $\alpha_{\mathrm{aniso}}$ vs. coupling constant $D$ (black diamonds). The red line is a quadratic fit, indicative of $\alpha_{\mathrm{aniso}} \propto D^2$.}
	\label{fig:aniso_alpha_M0}
\end{figure}

Next, we investigate the dependence of the Gilbert parameter $\alpha_{\mathrm{aniso}}$ on the bath coupling. Fig.~\ref{fig:aniso_alpha_M0} shows that $\alpha_{\mathrm{aniso}}$ increases quadratically with the electron-phonon coupling strength $D$. 

Since Fig.~\ref{fig:M0_T12} establishes that the spin-dephasing rate $1/T_2$ for the fast dynamics discussed in the previous sections, is proportional to $D^2$, we find $\alpha_{\mathrm{aniso}} \propto 1/T_2$. 
We briefly compare these trends to two earlier calculations of Gilbert damping that employ $p$-$d$ models and assume phenomenological Bloch-type rates $1/T_2$ for the dephasing of the itinerant hole spins toward the field of the localized moments. In contrast to the present paper, the localized spins experience the anisotropy fields. Chovan and Perakis~\cite{Perakis} derive a Gilbert equation for the dephasing of the localized spins toward the anisotropy axis, assuming that the hole spin follows the field $\vec{H}_{\mathrm{loc}}$ of the localized spins almost adiabatically. Tserkovnyak et al.~\cite{Tserkovnyak_APL04} extract a Gilbert parameter from spin susceptibilities. The resulting dependence of the Gilbert parameter $\alpha_{\mathrm{aniso}}$ on $1/T_2$ in both approaches is in qualitative accordance and exhibits two different regimes. In the the low spin-flip regime, where $1/T_2$ is small in comparison to the $p$-$d$ exchange interaction a linear increase of $\alpha_{\mathrm{aniso}}$ with $1/T_2$ is found, as is the case in our calculations with microscopic dephasing terms. If the relaxation rate is larger than the $p$-$d$ dynamics, $\alpha_{\mathrm{aniso}}$ decreases again. Due to the restriction~\eqref{markov_condition} of the Boltzmann scattering integral to low spin-flip rates, the present Markovian calculations cannot be pushed into this regime.

\begin{figure}[tb]
	\centering
		\includegraphics[width=0.49\textwidth]{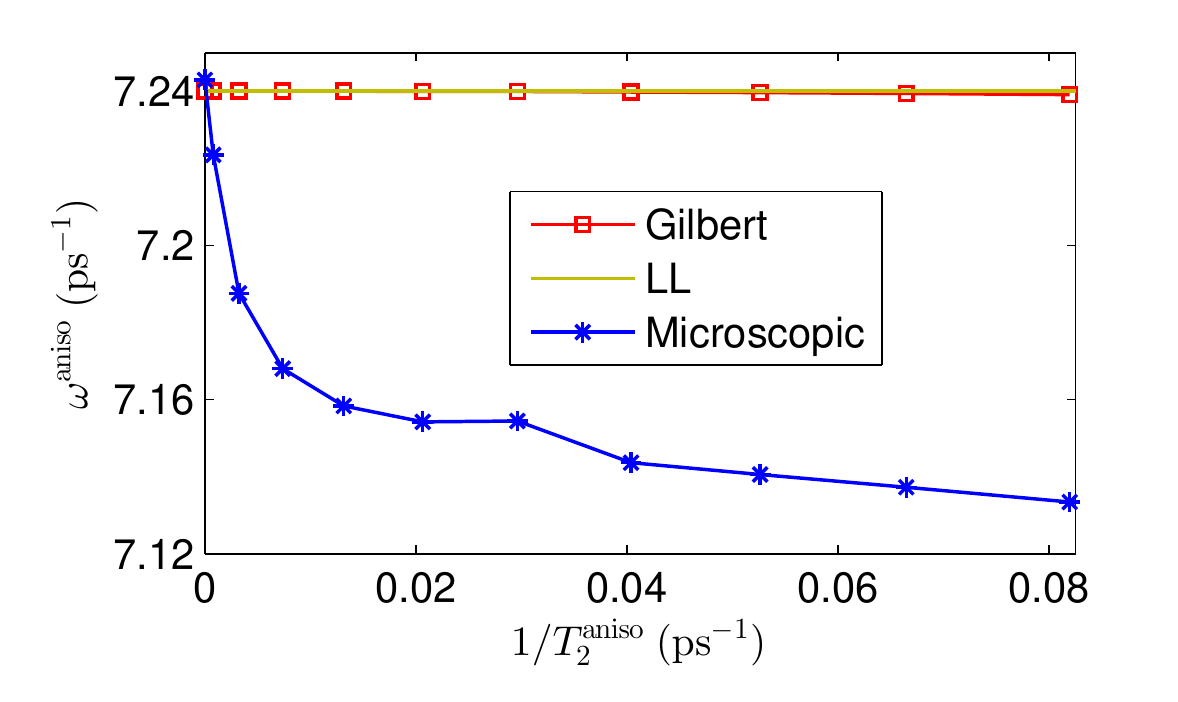}
	\caption{Precession frequency of the localized spin around the anisotropy field vs. Bloch decay time $1/T^{\text{aniso}}_2$.}
	\label{fig:aniso_Larmor}
\end{figure}

Even though the anisotropy field $\vec{H}_{\mathrm{aniso}}$ is not coupled to the localized spin $\vec{S}$ directly, both spins precess around the $z$ direction with frequency $\omega_{\mathrm{L}}^{\mathrm{aniso}}$. In analogy to Sec.~\ref{varying_larmor} we study now the influence of the damping process on the precession of the localized spin around the anisotropy axis and compare it to the behavior of Landau-Lifshitz and Gilbert dynamics. Fig.~\ref{fig:aniso_Larmor} reveals a similar behavior of the precession frequency as a function of the damping rate $1/T^{\text{aniso}}_2$ as in the isotropic case. The microscopic calculation predicts a distinct drop of the Larmor frequency $\omega_{\mathrm{L}}^{\mathrm{aniso}}$ for a range of dephasing rates where the precession frequency is unchanged according to the Gilbert and Landau-Lifshitz damping models. Although Gilbert damping eventually leads to a change in precession frequency for larger damping, this result shows a qualitative difference between the microscopic and the phenomenological calculations.

\section{Conclusion and Outlook \label{conclusion}}

In this paper, we investigated  a microscopic description of dephasing processes due to spin-orbit coupling and electron-phonon scattering in a mean-field kinetic exchange model. We first analyzed how spin-dependent carrier dynamics can be described by Boltzmann scattering integrals, which leads to Elliott-Yafet type relaxation processes. This is only possible for dephasing rates small compared to the Larmor frequency, see Eq.~\eqref{markov_condition}. The microscopic calculation always yielded Bloch times $2 T_1 = T_2$ for low excitation angles as it should be due to the conservation of the absolute value of the magnetization. A small decrease of the effective precession frequency occurs with increasing damping rate, which is a fundamental difference to the Landau-Lifshitz description and exceeds the change predicted by the Gilbert equation in this regime. 

We modeled two dephasing scenarios. First, a relaxation process between both spin sub systems was studied. Here, the different spins precess around the mean-field of the other system. In particular, for large excitation angles we found a decrease of the magnetization during the precessional motion without heating and a slow remagnetization. This scenario is somewhat in between typical small angle-relaxation, for which the modulus of the magnetization is constant and which is well described by Gilbert and Landau-Lifshitz damping, and collinear de/remagnetization dynamics. Also, we find important deviations from a pure Bloch-like behavior.

The second scenario deals with the relaxation of the magnetization toward a magnetic anisotropy field experienced by the itinerant carrier spins for small excitation angles. The resulting Gilbert parameter $\alpha_{\mathrm{aniso}}$ is independent of the static anisotropy field. The relaxation of the localized moments occurs only indirectly as a carrier-meditated relaxation via their coupling to the time dependent mean-field of the itinerant spin. 

To draw a meaningful comparison with Landau-Lifshitz and Gilbert dynamics we restricted ourselves throughout the entire paper to a regime where the electronic temperature is equal to the lattice temperature $T_{\mathrm{ph}}$ at all times. In general our microscopic theory is also capable of modeling heat induced de- and remagnetization processes. We intend to compare microscopic simulations of hot electron dynamics in this model, including scattering processes between both types of spin, with phenomenological approaches such as the Landau-Lifshitz-Bloch (LLB) equation or the self-consistent Bloch equation (SCB)~\cite{Zhang_2013}.    

We finally mention that we derived relation~\eqref{alpha_T2_relation} connecting the Bloch dephasing time $T_2$ and the Gilbert damping parameter $\alpha$. Despite its simplicity and obvious usefulness, we were not able to find a published account of this relation.



\begin{thebibliography}{99}
%
\bibitem{LandauLifshitz} L.~Landau and E.~Lifshitz, Phys. Z. Sowj. \textbf{8}, 153 (1935).

\bibitem{Gilbert} T.~L.~Gilbert, IEEE Trans. Magnetics, \textbf{40}, 6 (2004).

\bibitem{Mills_LLG} D.~L.~Mills and R.~Arias \textit{The damping of spin motions in ultrathin films: Is the Landau-Lifschitz-Gilbert phenomenology applicable?} Invited paper presented at the VII Latin American Workshop on
Magnetism, Magnetic Materials and their Applications, Renaca, Chile, 2005.

\bibitem{Garanin_PhysicaA91} D. A. Garanin, Physica A~\textbf{172}, 470 (1991).

\bibitem{Gerasimenko} V.~V.~Andreev and V.~I.~Gerasimenko, Sov. Phys. JETP \textbf{35}, 846 (1959).

\bibitem{Bauer} A.~Brataas, A.~Tserkovnyak, and G.~E.~W.~Bauer, \prl~\textbf{101}, 037207 (2008).

\bibitem{Saslow} W.~M.~Saslow, J. Appl. Phys.~\textbf{105}, 07D315 (2009).

\bibitem{Smith} N.~Smith, \prb~\textbf{78}, 216401 (2008).

\bibitem{Zhang_2009} S. Zhang and S.-L.~Zhang, \prl~\textbf{102}, 086601 (2009).

\bibitem{Nowak_Switching} S.~Wienholdt, D.~Hinzke, K.~Carva. P.~M.~Oppeneer and U.~Nowak, \prb~\textbf{88}, 020406(R) (2013).

\bibitem{Nowak_Ultrafast} U.~Atxitia, O.~Chubykalo-Fesenko, R.~W.~Chantrell, U.~Nowak and A.~Rebei, \prl~\textbf{102}, 057203 (2009).

\bibitem{Zhang_2004} Z.~Li and S.~Zhang, \prb~\textbf{69}, 134416 (2004).

\bibitem{Fesenko_LLB} O.~Chubykalo-Fesenko, \apl~\textbf{91}, 232507 (2007).

\bibitem{Garanin_PRB97} D.~A.~Garanin, \prb~\textbf{55}, 3050-3057 (1997).

\bibitem{Fesenko_Ultrafast} U.~Atxitia, and O.~Chubykalo-Fesenko, \prb~\textbf{84}, 144414 (2011).

\bibitem{Garate_PRB2009} I.~Garate and A.~MacDonald, \prb~\textbf{79}, 064403 (2009)

\bibitem{Gilmore_PRB2011} K.~Gilmore, I.~Garate, A.~H.~MacDonald, and M.~D.~Stiles, \prb~\textbf{84}, 224412 (2011)

\bibitem{Sinova} J.~Sinova, T.~Jungwirth, X.~Liu, Y.~Sasaki, J.~K.~Furdyna, W.~A.~Atkinson, and A.~H.~MacDonald, \prb~\textbf{69}, 085209 (2004).

\bibitem{Shen_2010} K.~Shen, G.~Tatara, and M.~W.~Wu, \prb~\textbf{81}, 193201 (2010)

\bibitem{Shen_2012} K.~Shen and M.~W.~Wu, \prb~\textbf{85}, 075206 (2012)

\bibitem{MacDonald_Review} T.~Jungwirth, J.~Sinova, J.~Masek, J.~Kucera and A.~H.~MacDonald, \rmp~\textbf{78}, 809 (2006).

\bibitem{Dietl_Review} T.~Dietl, Nature Mat.~\textbf{9}, 2898 (2010).

\bibitem{MacDonald_Curie} J.~Koenig, H.-H.~Lin and A.~H.~MacDonald, \prl~\textbf{84}, 5628 (2000).

\bibitem{MacDonald_SpinWave} J.~K\"{o}nig, T.~Jungwirth and A.~H.~MacDonald, \prb~\textbf{64}, 184423 (2001).

\bibitem{Rasing} A. Kirilyuk, A. V. Kimel, and T. Rasing, \rmp~\textbf{82}, 2731 (2010).

\bibitem{Alebrand} S. Mangin, M. Gottwald, C-H. Lambert, D. Steil, V. Uhlír, L. Pang, M. Hehn, S. Alebrand, M. Cinchetti, G. Malinowski, Y. Fainman, M. Aeschlimann and E. E. Fullerton, Nature Materials~13, 286 (2014).

\bibitem{Bloch} F.~Bloch, Phys.\ Rev.~\textbf{70}, 460 (1946)

\bibitem{Bloembergen} N.~Bloembergen, Phys.\ Rev.~\textbf{78}, 572 (1950)

\bibitem{Vonsovskii} S.~V.~Vonsovskii, \textit{Ferromagnetic Resonance} (Pergamon, Oxford, 1966).

\bibitem{Stancil} D.~D.~Stancil and A. Prabhakar, \textit{Spin Waves}, (Springer, New York, 2009).

\bibitem{Slichter} D.~Pines and C.~Slichter, Phys.\ Rev.~\textbf{100}, 1014 (1955).

\bibitem{Wu} C.~L{\"{u}}, J.~L.~Cheng, M.~W.~Wu and I.~C.~da Cunha Lima, Phys.\ Lett.~A~\textbf{365}, 501 (2007).

\bibitem{Leontiadou_2011} M. Leontiadou, K.~L.~Litvinenko, A.~M.~Gilbertson, C.~R.~Pidgeon, W.~R.~Branford, L.~F.~Cohen, M.~Fearn, T.~Ashley, M.~T.~Emeny, B.~N.~Murdin and S.~K.~Clowes, J. Phys.: Condens.\ Matter \textbf{23}, 035801 (2011).

\bibitem{HaugKoch} H. Haug and S. W. Koch, \textit{Quantum Theory of the Optical and Electronic Properties of Semiconductors}, 4. ed (World Scientific, Singapore, 2003).

\bibitem{Jauho_Phonon} K.~Kaasbjerg, K.~S.~Thygesen and A.-P.~Jauho, \prb~\textbf{87}, 235312 (2013).

\bibitem{Raichev} F.~T.~Vasko and O.~E.~Raichev, \textit{Quantum Kinetic Theory and Applications}, (Springer, New York, 2005).

\bibitem{SchneiderChowKoch} H. C. Schneider, W. W. Chow, and S. W. Koch, \prb~\textbf{70}, 235308 (2004).





\bibitem{Perakis} J.~Chovan and I.~E.~Perakis, \prb~\textbf{77}, 085321 (2008).

\bibitem{Tserkovnyak_APL04} Y. Tserkovnyak, G. A. Fiete, and B. I. Halperin, \apl~\textbf{84}, 5234 (2004).

\bibitem{Zhang_2013} L. Xu and S. Zhang, J. Appl.\ Phys.~\textbf{113}, 163911 (2013).

%
\end{thebibliography}
\end{document}